\documentclass[pdflatex,sn-basic,natbib,iicol]{sn-jnl}% Springer Nature Basic Reference Style 
\makeatletter
\let\orcid\@undefined
\let\orcidlogo\@undefined
\makeatother
\usepackage[authoryear]{natbib}
\usepackage{graphicx}%
\usepackage{multirow}%
\usepackage{amsmath,amssymb,amsfonts}%
\usepackage{amsthm}%
\usepackage{mathrsfs}%
\usepackage{bm}%
\usepackage[title]{appendix}%
\usepackage{xcolor}%
\usepackage{textcomp}%
\usepackage{manyfoot}%
\usepackage{booktabs}%
\usepackage{algorithm}%
\usepackage{algorithmicx}%
\usepackage{algpseudocode}%
\usepackage{listings}%
\usepackage{orcidlink}% TikZ-based ORCID logo (sn-jnl.cls's \orcid uses Orcidlogo.eps which is not shipped)
\usepackage{float}% provides [H] placement specifier for figures
\AtBeginDocument{}% sn-jnl.cls's \burl breaks on underscores under pdflatex; route to hyperref's \url
\theoremstyle{thmstyleone}%
\theoremstyle{thmstyletwo}%

\theoremstyle{thmstylethree}%

\begin{document}

\title[Multiple mechanisms of rhythm switching in RNNs]{Multiple mechanisms of rhythm switching in recurrent neural networks with adaptive time constants}

\author*[1]{\fnm{Yutaka} \sur{Yamaguti}\,\orcidlink{0000-0002-1739-3387}}\email{y-yamaguchi@fit.ac.jp}

\author[2]{\fnm{Shota} \sur{Nakamura}}

\affil*[1]{\orgdiv{Faculty of Information Engineering}, \orgname{Fukuoka Institute of Technology}, \orgaddress{\street{3-30-1 Wajiro-Higashi, Higashi-ku}, \city{Fukuoka}, \postcode{811-0295}, \state{Fukuoka}, \country{Japan}}}

\affil[2]{\orgdiv{Graduate School of Engineering}, \orgname{Fukuoka Institute of Technology}, \orgaddress{\street{3-30-1 Wajiro-Higashi, Higashi-ku}, \city{Fukuoka}, \postcode{811-0295}, \state{Fukuoka}, \country{Japan}}}

\abstract{Although recurrent neural networks (RNNs) trained on cognitive tasks have become a widely used framework for studying neural computation, the internal mechanisms by which RNNs switch between rhythms across multiple frequency bands, and how these mechanisms relate to neuronal time constants, have not been systematically analyzed. 
We trained leaky integrator RNNs with neuron-specific learnable time constants on a four-band (theta, alpha, beta, gamma) rhythm-switching task and analyzed 20 independently trained networks.
Whereas low-frequency rhythms were produced by distributed participation of many neurons, high-frequency rhythms were dominated by a small subpopulation of short-time-constant neurons, and the negative correlation between time constant and matched-mode amplitude strengthened monotonically with frequency. 
Rhythm switching was supported by multiple coexisting mechanisms: turnover of the active subpopulation, network-wide baseline shifts that reposition the operating point near distinct unstable fixed points, and inter-neuronal phase reorganization that selectively cancels or supports band components in the population output.
Analysis of the learned recurrent connectivity linked these mechanisms to network structure: the short-time-constant neurons formed a more strongly interconnected module, and the baseline shift retuned the oscillation frequency through gain modulation of the recurrent interactions.
The mechanism deployed for each mode pair varied across training runs, exposing a degeneracy of learned solutions.
These findings parallel the coexistence of rhythm-specific and multi-rhythm interneurons reported in biological circuits and provide a candidate framework for interpreting frequency-band-specific functional differentiation in neural systems.
}


\maketitle

\section{Introduction}\label{sec:introduction}

Analyzing the dynamic computational structures that emerge within recurrent neural networks (RNNs) through task learning has become a central methodology in computational neuroscience.
 \cite{sussillo_barak2013} applied fixed-point analysis to trained RNNs and characterized the computational mechanisms of discrete decision-making tasks in terms of attractor geometry. 
 \citet{maheswaranathan2019} showed through low-dimensional dynamics analysis that learned dynamical systems are efficiently organized on low-dimensional manifolds. \cite{yang2019task} further demonstrated, using a multi-task learning framework, that functional modular structures spontaneously emerge within RNNs. 
Separately, Yamaguti and colleagues \citep{yamaguti2021functional, tomoda2026emergence} systematically investigated functional differentiation in RNNs, showing that it arises depending on task properties and constraints and that information-theoretic objectives can be used to induce it. 
Collectively, these studies have established that trained RNNs are not ``black boxes'' but acquire interpretable internal structures shaped by the computational demands of the task. 

Most tasks examined in the above studies involve either steady states and slow temporal dynamics, such as decision-making and working memory, or relatively simple rhythmic oscillations. 
Rhythmic oscillation is an important class of computation that RNNs can learn, and RNNs are capable of generating complex rhythmic patterns \citep{sussillo2009}. 
\citet{doyayoshizawa1989memorizing} presented a pioneering framework for training RNNs on periodic patterns with multiple time scales using continuous-time backpropagation. 
\cite{sussillo2009} introduced FORCE learning, enabling reservoir-type RNNs to generate arbitrary temporal patterns. 
\citet{sussillo_barak2013} showed, in a sinusoidal generation task in which input specifies the frequency, that the oscillation frequency is governed by an operating point on a low-dimensional manifold. 
More recently, \cite{zemlianova2024} trained excitatory--inhibitory RNNs on rhythmic timing tasks and elucidated the oscillation generation mechanism using low-dimensional models, and \cite{pals2024} reported that phase-locked limit cycles emerge in a combined oscillation and working-memory task. 
However, all of these studies focused on the macroscopic dynamics of oscillation generation (low-dimensional manifolds, limit cycles, etc.), and none has systematically analyzed how individual neurons functionally differentiate across frequency bands, specifically, which neurons serve which bands and how neuronal populations reorganize when the rhythm switches.

In biological neural circuits, individual neurons participate in multiple rhythms in diverse ways. 
A comprehensive review by \citet{klausberger2008} on hippocampal CA1 interneurons described cell types such as O-LM cells that fire specifically during the $\theta$ rhythm, alongside basket cells and bistratified cells that participate in both $\theta$ and $\gamma$ rhythms and shift their firing phase when the dominant rhythm changes. 
Biological circuits thus harbor both rhythm-specific specialists and generalists that flexibly engage in multiple rhythms. 
Whether such diverse functional differentiation can emerge spontaneously within RNNs through supervised learning has not yet been systematically examined.

The relationship between neuronal time constants and functional differentiation in RNNs has long attracted attention. 
\citet{yamashita2008} proposed the multiple timescale RNN, featuring two classes of context units with fast and slow time constants. 
In humanoid robot motor sequence tasks, a functional hierarchy self-organized without explicit hierarchical architecture such that fast units encoded reusable motor primitives while slow units encoded the sequential ordering of those primitives. 
Although this pioneering study established that differences in neuronal time constants can drive functional differentiation, the time constants were fixed at two discrete values \textit{a priori} and were not themselves learned.

More recently, learnable time constant diversity has been shown to enhance the efficiency and robustness of RNNs. 
\cite{perez-nieves2021} demonstrated that time constant heterogeneity improves learning robustness in spiking neural networks and reported that the learned distribution qualitatively matches experimental data. 
\cite{quax2020} showed that a two-layer adaptive-time-constant RNN self-organizes a temporal-scale hierarchy across layers through training. 
\cite{stern2023} theoretically analyzed how multiple time scales emerge from heterogeneous neuronal populations and pointed out that single-neuron time constant diversity alone does not necessarily produce time-scale separation under recurrent dynamics. 
\cite{nuttida2025} reported that noise during training promotes time constant diversification and improves working memory performance.

These findings establish that time constant differences can produce functional differentiation \citep{yamashita2008} and that learnable time constants enrich RNN representational capacity \citep{perez-nieves2021, quax2020}. 
However, how the learned continuous distribution of time constants maps onto frequency-band-specific functional differentiation of neuronal populations has not been directly analyzed. 
In particular, what population-level structures arise from time constant diversity in rhythm-switching tasks involving multiple time scales remains an open question.

In this study, we address these open questions. 
Specifically, in RNNs trained to switch between rhythms spanning multiple frequency bands, we systematically investigate (i)~how the learned continuous distribution of time constants corresponds to frequency-band-specific functional differentiation, and (ii)~what internal mechanisms underlie transitions between rhythms. 
We employ a leaky integrator RNN in which the leak rate of each neuron is a learnable parameter and design a differentiable loss function based on spectral band power ratios. 
The network is trained on a rhythm-switching task across four frequency bands ($\theta$: 4--7\,Hz, $\alpha$: 8--13\,Hz, $\beta$: 14--29\,Hz, $\gamma$: 30--50\,Hz). 
We further introduce an amplitude-weighted synchronization measure (absolute sync strength) that simultaneously captures amplitude and phase synchrony, enabling systematic analysis of the functional differentiation that emerges internally. 
Analysis of multiple independently trained models reveals the following main findings.

First, frequency-dependent heterogeneity emerges in the amplitude distribution: low-frequency rhythms ($\theta$, $\alpha$) are generated through a distributed mechanism in which many neurons contribute uniformly, whereas high-frequency rhythms ($\beta$, $\gamma$) rely on a concentrated mechanism in which amplitude is dominated by a small number of high-amplitude neurons. 
Second, this functional differentiation systematically corresponds to the learned time constants: neurons with shorter time constants are selectively recruited for higher-frequency bands, while this selectivity weakens for lower frequencies. 
Third, for the same task and loss function, robust and variable aspects coexist in the learned solutions: the independence of the $\theta$-rhythm neuronal population is robust across all training runs, whereas population-sharing patterns among $\alpha$, $\beta$, and $\gamma$ exhibit diverse configurations across runs.

Taken together, these findings indicate that the internal mechanisms of rhythm switching cannot be explained by a single principle, and at least three distinct mechanisms act in combination: 
(A) population turnover---low- and high-frequency oscillations engage different subpopulations (many neurons for low-frequency rhythms, a small dominant subpopulation for high-frequency rhythms), so that rhythm switching is accompanied by turnover of the active population; 
(B) baseline shift---without substantial turnover of the active population, a shift in mean activity level repositions the operating point of the dynamical system; 
(C) phase-interference-driven frequency shift---individual neurons maintain their oscillations, but reorganization of inter-neuronal phase coherence shifts the spectral peak of the population signal. 
This three-mechanism framework qualitatively parallels the coexistence of rhythm-specific neurons and multi-rhythm-participating neurons observed in biological neural circuits \citep{klausberger2008}, suggesting that RNNs can spontaneously acquire analogous functional organization in response to task demands.

The remainder of this paper is organized as follows. Sect.~\ref{sec:methods} describes the model formulation, task design, loss function, and analysis methods. Sect.~\ref{sec:results} reports the discovery of frequency-dependent functional differentiation, its correspondence with time constants, inter-mode population similarity and its variability across training runs, baseline shift as a substrate for rhythm switching, the causal role of fast neurons revealed by silencing experiments, phase-interference-mediated control of band power, the structural and dynamical basis of these mechanisms in the recurrent connectivity and gain modulation, and the robustness of the results to the time-constant initialization.
Sect.~\ref{sec:discussion} discusses comparisons with prior work on time constant diversity, the significance of solution diversity, and implications for neuroscience.

\section{Methods}\label{sec:methods}

\subsection{Model}\label{sec:model}

We build on the leaky integrator RNN, a standard model in computational neuroscience, and extend it by treating the leak rate of each neuron (the reciprocal of its time constant) as a learnable parameter.
The network consists of $N = 100$ neurons whose internal state $\bm{x}(t) \in \mathbb{R}^N$ obeys the following discrete-time dynamics:
\begin{equation}
\begin{split}
  \bm{x}(t+1) = {} & (\bm{1} - \bm{\lambda}) \odot \bm{x}(t) \\
  & + \bm{\lambda} \odot \big( W_{\mathrm{rec}}\, \bm{r}(t) + W_{\mathrm{in}}\, \bm{u}(t) \\
  & \qquad + \bm{b} + \sigma_\xi \bm{\xi}(t) \big),
\end{split}
\label{eq:dynamics}
\end{equation}
\begin{equation}
  \bm{r}(t) = \tanh(\bm{x}(t)),
  \label{eq:activation}
\end{equation}
where $\bm{r}(t) \in \mathbb{R}^N$ is the firing-rate vector,
$\bm{u}(t) \in \mathbb{R}^K$ is the task input ($K = 4$ channels),
$\bm{\xi}(t) \sim \mathcal{N}(\bm{0}, \bm{I}_N)$ is independent Gaussian noise,
$\sigma_\xi = 0.1$ is the noise intensity,
$\bm{b} \in \mathbb{R}^N$ is the bias vector,
and $\odot$ denotes the Hadamard (element-wise) product.

The vector $\bm{\lambda} = (\lambda_1, \ldots, \lambda_N)^{\top}$ contains neuron-specific leak-rate parameters that are learned jointly with the other weight parameters under the constraint $\lambda_i \geq 0$.
The effective time constant of neuron $i$ is defined as $\tau_i = 1/\lambda_i$ (in units of time steps).
All leak rates are initialized uniformly to $\lambda_i = 0.02$ ($\tau_i = 50$ steps, i.e.\ 50\,ms at a sampling frequency $F_s = 1000$\,Hz); the robustness of the results to this choice is examined in Sect.~\ref{sec:reinit_results}.

Weight matrices are initialized as follows:
the recurrent weight matrix $W_{\mathrm{rec}} \in \mathbb{R}^{N \times N}$ is drawn independently from $\mathcal{N}(0, 1/N)$;
the input weight matrix $W_{\mathrm{in}} \in \mathbb{R}^{N \times K}$ is initialized with the Xavier uniform distribution;
and the bias $\bm{b}$ is initialized to zero.

The population-averaged output
\begin{equation}
  z(t) = \frac{1}{N} \sum_{i=1}^{N} r_i(t)
  \label{eq:readout}
\end{equation}
serves as the signal to be evaluated by the task.
The readout weights are fixed (a simple average over all neurons), so that training does not alter the relative contribution of individual neurons through the readout layer.
This design ensures that the properties of the population output $z(t)$ are determined solely by the internal dynamics of the RNN.

\subsection{Task}\label{sec:task}

The task requires the network to switch the rhythm (oscillation frequency band) of its population output $z(t)$ in response to input pulses.
We define $K = 4$ rhythm types:
$\theta$ (4--7\,Hz), $\alpha$ (8--13\,Hz), $\beta$ (14--29\,Hz), and $\gamma$ (30--50\,Hz).

Each trial consists of a waiting period followed by $M = 3$ rhythm intervals.
During the waiting period ($T_{\mathrm{wait}} = 100$ steps) at the beginning of the trial, the input is zero.
Each rhythm interval $m \in \{1, \ldots, M\}$ is assigned one of the $K$ rhythms at random ($k_m \in \{1, \ldots, K\}$), and a one-hot pulse $\bm{o}_m \in \{0,1\}^K$ corresponding to the selected rhythm is fed to the network as the input $\bm{u}(t) = \bm{o}_m$ (Eq.~\ref{eq:dynamics}) for $T_{\mathrm{pulse}} = 100$ steps at the onset of that interval:
\begin{equation}
  (\bm{o}_m)_k = \begin{cases} 1 & \text{if } k = k_m, \\ 0 & \text{otherwise.} \end{cases}
\end{equation}
For the remainder of each interval, the input is zero.
The one-hot pulse serves as a discrete contextual command that selects the target rhythm, rather than as an oscillatory signal to be tracked; we discuss the motivation for this design choice in Sect.~\ref{sec:input_discussion}.

\begin{figure*}[t]
  \centering
  \includegraphics[width=\textwidth]{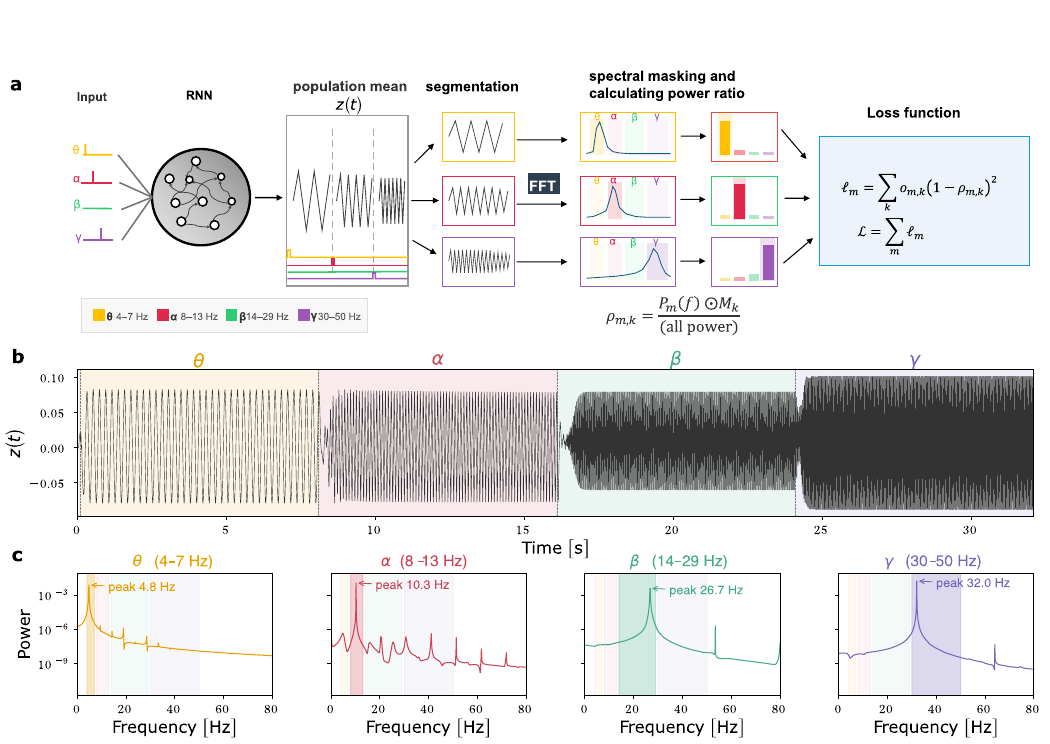}
  \caption{Task overview and training results.
  (a) Task schematic. The network switches among four rhythm modes in response to one-hot pulse inputs.
  (b) Time series of the population output $z(t)$ for a representative trial.
  (c) Power spectra of each rhythm interval (log scale; all four panels share a common y-axis range).}
  \label{fig:task_overview}
\end{figure*}

Each interval has a duration of $T_{\mathrm{int}} = 4500$ steps (4.5\,s) with the first $T_{\mathrm{tr}} = 500$ steps (0.5\,s) treated as a transition period and excluded from the loss evaluation.
The effective signal evaluation length is thus $T_{\mathrm{int}} - T_{\mathrm{tr}} = 4000$ steps (4.0\,s).
The sampling frequency is $F_s = 1000$\,Hz.
Within each mini-batch, the rhythm sequences are generated independently at random.
Hereafter, we refer to the frequency band $k_m$ assigned to each interval $m$ as the rhythm mode and index it by the interval number $m$.
Figure~\ref{fig:task_overview}a shows the task schematic.

\subsection{Loss function}\label{sec:loss}

The loss function evaluates the degree to which the power spectrum of the population output $z(t)$ in each interval is concentrated within the frequency band of the designated rhythm.

Let $t_m$ denote the start time of interval $m$.
For the output signal $z_m(t)$ of interval $m$, taken over the analysis window after the transition period ($t \in [t_m + T_{\mathrm{tr}},\; t_m + T_{\mathrm{int}})$), we compute the power spectrum via the FFT:
\begin{equation}
  P_m(f) = \frac{|Z_m(f)|^2}{F_s \cdot T},
\end{equation}
where $Z_m(f)$ is the FFT of $z_m(t)$ (excluding the DC component) and $T$ is the signal length.

Using a binary mask $\bm{M}_k \in \{0,1\}^{N_f}$ that selects the frequency bins of rhythm $k$ (i.e.\ $[f_k^{\mathrm{low}},\, f_k^{\mathrm{high}})$), we compute the band power ratio:
\begin{equation}
  \rho_{m,k} = \frac{\sum_{n} P_m(f_n) \cdot (\bm{M}_k)_n}{\sum_{n} P_m(f_n)}.
\end{equation}
The per-interval loss is defined so that the power ratio $\rho_{m,k_m}$ of the designated rhythm $k_m$ approaches~1:
\begin{equation}
  \ell_m = \sum_{k=1}^{K} (\bm{o}_m)_k \left( 1 - \rho_{m,k} \right)^2.
\end{equation}
The overall loss is averaged over all samples and intervals within a mini-batch (batch size $B$):
\begin{equation}
  \mathcal{L} = \frac{1}{B \cdot M} \sum_{b=1}^{B} \sum_{m=1}^{M} \ell_{m,b}.
  \label{eq:loss}
\end{equation}

This loss function evaluates only the power ratio (the relative spectral composition) and imposes no constraint on absolute amplitude or on the dynamics of individual neurons.
Consequently, the internal mechanism by which the network realizes rhythm switching is left entirely free to be determined by learning.

\subsection{Training}\label{sec:training}

We use Adam \citep{kingma2015adam} with a learning rate $\eta = 10^{-4}$ and global gradient clipping $\|\nabla\|_2 \leq 1$ \citep{pascanu2013difficulty}.
The batch size is $B = 50$.
Training runs for at most 50{,}000 iterations and is terminated early when $\mathcal{L} < 10^{-6}$.

We performed $n = 20$ independent training runs with different random seeds.

\subsection{Analysis methods}\label{sec:analysis}

To analyze the internal dynamics of the trained networks, we apply the methods described below.
For the analysis, we use an evaluation protocol in which each rhythm band is presented once in sequence: we set the number of intervals to $M = K$ and assign rhythm $k_m = m$ to interval $m$ (i.e.\ $m = 1$: $\theta$, $m = 2$: $\alpha$, $m = 3$: $\beta$, $m = 4$: $\gamma$).
Each interval is extended to $T_{\mathrm{int}}^{\mathrm{eval}} = 8000$ steps (8.0\,s) to obtain stable spectral and synchronization estimates. The initial $T_{\mathrm{tr}} = 500$ steps are excluded as a transition period, as during training, and the network is run noise-free ($\sigma_\xi = 0$).
Under this setting, the interval index $m$ simultaneously serves as the rhythm mode (Sect.~\ref{sec:task}) and band index, so we write $k_m$ simply as $m$.%

\subsubsection{Bandpass filtering and amplitude extraction}\label{sec:bandpass}

We apply a fourth-order Butterworth bandpass filter corresponding to each frequency band $k \in \{1, \ldots, K\}$ to obtain a band-limited signal $\tilde{r}_i^{(k)}(t)$ for the firing rate $r_i(t)$ of each neuron.
We then apply the Hilbert transform to construct the analytic signal
\begin{equation}
  \zeta_i^{(k)}(t) = \tilde{r}_i^{(k)}(t) + \mathrm{i}\,\mathcal{H}[\tilde{r}_i^{(k)}](t) = A_i^{(k)}(t)\,e^{\mathrm{i}\phi_i^{(k)}(t)},
\end{equation}
from which we extract the instantaneous amplitude $A_i^{(k)}(t) \geq 0$ and instantaneous phase $\phi_i^{(k)}(t) \in (-\pi,\pi]$.

The band amplitude during rhythm mode $m$ (over the analysis window $\mathcal{T}_m$ that excludes the transition period $T_{\mathrm{tr}}$) is defined as the temporal mean of the Hilbert envelope:
\begin{equation}
  \bar{A}_i^{(m,k)} = \left\langle A_i^{(k)}(t) \right\rangle_{t \in \mathcal{T}_m}.
  \label{eq:band_amplitude}
\end{equation}
In particular, the amplitude for the band that matches mode $m$, i.e.\ $\bar{A}_i^{(m,m)}$, is termed the matched-mode amplitude.

\subsubsection{Amplitude distribution statistics}\label{sec:amp_stats}

The distribution of matched-mode amplitudes across the neuronal population is characterized by its skewness and Gini coefficient, which together quantify the degree to which participation in oscillations is uniformly distributed across neurons or concentrated in a small subpopulation.

\subsubsection{Absolute sync strength and sync degree}\label{sec:sync_strength}

To quantify the synchronization between a pair of neurons $(i,j)$ in mode $m$, we define the \textbf{absolute sync strength} $S_{ij}^{(m)}$ from the instantaneous amplitudes and phases of the matched band as
\begin{equation}
\begin{split}
  S_{ij}^{(m)} & = \left| \left\langle A_i^{(m)}(t)\,A_j^{(m)}(t)\,e^{\mathrm{i}\Delta\phi_{ij}^{(m)}(t)} \right\rangle_{t \in \mathcal{T}_m} \right|, \\
  \Delta\phi_{ij}^{(m)}(t) & = \phi_i^{(m)}(t) - \phi_j^{(m)}(t),
\end{split}
\label{eq:abs_sync}
\end{equation}
where $\langle\cdot\rangle_{t\in\mathcal{T}_m}$ denotes the temporal mean over the analysis window $\mathcal{T}_m$.
When the mode is clear from context, we omit the superscript $(m)$ and write $S_{ij}$, $A_i(t)$, $\phi_i(t)$.

The standard phase locking value (PLV) $R_{ij} = |\langle e^{\mathrm{i}\Delta\phi_{ij}(t)}\rangle|$ and its variants evaluate only the phase relationship and ignore amplitude.
Consequently, neurons that barely oscillate can yield a high PLV through accidental phase alignment, an artifact that is particularly pronounced for bandpass-filtered signals and that we indeed observed in the present data.
Because $S_{ij}$ is not normalized by a denominator, a large value of $S_{ij}$ requires both (1)~that both neurons have substantial amplitude and (2)~that their phase difference is stable.
This property enables $S_{ij}$ to capture synchronization only among neurons that genuinely participate in oscillations.

To summarize the pairwise structure at the single-neuron level, we define the \textbf{sync degree} of neuron $i$ as the row sum of $S^{(m)}_{ij}$:
\begin{equation}
  d^{(m)}_i = \sum_{j \neq i} S^{(m)}_{ij},
  \label{eq:sync_degree}
\end{equation}
which corresponds to the degree in a weighted graph and quantifies how strongly neuron $i$ participates in the collective synchronous activity of the network.

\subsubsection{Correlation with time constants}\label{sec:tau_corr}

We evaluate the relationship between the learned time constants $\tau_i = 1/\lambda_i$ and the activity level of each neuron in each mode using two measures.

First, we compute the Spearman rank correlation between the matched-mode amplitude $\bar{A}_i^{(m,m)}$ and $\tau_i$:
\begin{equation}
  \rho_s^{\mathrm{amp}}(m) = \mathrm{Spearman}\left(\bar{A}_i^{(m,m)},\; \tau_i\right).
  \label{eq:tau_amp_corr}
\end{equation}
This directly tests whether neurons with particular intrinsic time constants exhibit larger amplitudes in the corresponding rhythm mode.

Second, we compute the Spearman rank correlation between the sync degree $d_i$ (Eq.~\ref{eq:sync_degree}) and $\tau_i$:
\begin{equation}
  \rho_s^{\mathrm{sync}}(m) = \mathrm{Spearman}\left(d_i,\; \tau_i\right).
  \label{eq:tau_sync_corr}
\end{equation}
This evaluates the relationship between time constants and the role of each neuron in the synchronization network.

Together, these two correlations test how time constants map onto both the amplitude of individual neurons (single-unit level) and their participation in collective synchrony (network level).

\subsubsection{Inter-mode population similarity}\label{sec:intermode}

The similarity of neuronal population amplitude patterns across different rhythm modes is quantified using two measures.

First, we compute the Spearman rank correlation between the matched-mode amplitude vectors $\bm{\bar{A}}^{(m)} = (\bar{A}_1^{(m,m)}, \ldots, \bar{A}_N^{(m,m)})^{\top}$:
\begin{equation}
  \rho_{m_1 m_2}^{\mathrm{amp}} = \mathrm{Spearman}\left(\bm{\bar{A}}^{(m_1)},\; \bm{\bar{A}}^{(m_2)}\right).
  \label{eq:amp_corr}
\end{equation}

Second, we extract the top 20\% of neurons by sync degree in each mode as the active set $\mathcal{A}_m$ and evaluate the overlap between mode pairs using the Jaccard coefficient:
\begin{equation}
  J(m_1, m_2) = \frac{|\mathcal{A}_{m_1} \cap \mathcal{A}_{m_2}|}{|\mathcal{A}_{m_1} \cup \mathcal{A}_{m_2}|}.
\end{equation}

\subsubsection{Baseline analysis: fixed-point search and stability analysis}\label{sec:baseline}

To quantitatively test whether a shift in baseline activity level (mean firing rate) governs the oscillation frequency during the $\beta$--$\gamma$ rhythm transition, we perform the following three-stage analysis.
This analysis targets a training run in which the synchronized populations for $\beta$ and $\gamma$ overlap substantially (Jaccard coefficient $J(\beta,\gamma) = 0.667$).

\paragraph{Fixed-point search}
For each mode $m \in \{3, 4\}$ (i.e.\ $\beta$ and $\gamma$; see Sect.~\ref{sec:task}), we numerically search for fixed points $\bm{x}^*$ of the noise-free, input-free network dynamics (Eq.~\ref{eq:dynamics} with $\bm{u} = \bm{0}$, $\sigma_\xi = 0$).
Following \citet{sussillo_barak2013}, we search fixed points by minimizing the objective function
\begin{equation}
\begin{split}
  q(\bm{x}) & = \| f(\bm{x}) - \bm{x} \|^2, \\
  f(\bm{x}) & = (\bm{1} - \bm{\lambda}) \odot \bm{x} \\
  & \quad + \bm{\lambda} \odot \big( W_{\mathrm{rec}}\, \tanh(\bm{x}) + \bm{b} \big)
\end{split}
\label{eq:fp_objective}
\end{equation}
using L-BFGS-B method \citep{byrd1995} as implemented in SciPy library \citep{virtanen2020scipy}.
We use six initial conditions (the baseline itself plus five random perturbations with $\sigma = 0.01$) applied to the pre-activation state $\mathrm{arctanh}(\bar{\bm{r}}^{(m)})$ corresponding to the time-averaged activity of each mode, and cluster the converged solutions by Euclidean distance to identify distinct fixed points.
A fixed point is accepted when $q < 10^{-10}$.

\paragraph{Discrete-time Jacobian and stability analysis}
At the fixed point $\bm{x}^*$, we compute the Jacobian matrix of the discrete-time dynamics:
\begin{equation}
  J_{ij} = (1 - \lambda_i)\delta_{ij} + \lambda_i \, (1 - r_j^{*2}) \, (W_{\mathrm{rec}})_{ij},
  \label{eq:jacobian}
\end{equation}
where $r_j^* = \tanh(x_j^*)$.
Eigenvalues $\mu_k$ with $|\mu_k| > 1$ indicate unstable directions.
When $\mu_k$ is complex, the associated oscillation frequency is
\begin{equation}
  f_k = \frac{|\arg(\mu_k)|}{2\pi} \times F_s.
  \label{eq:eig_freq}
\end{equation}
We verify whether the frequency of the dominant unstable eigenvalue falls within the corresponding rhythm band.

\paragraph{Free-run simulation}
Starting from the pre-activation state $\mathrm{arctanh}(\bar{\bm{r}}^{(m)})$ corresponding to the mean activity of each mode, we run the network for 5000 steps without external input or noise and compute the power spectrum of the population output $z(t)$.
This directly tests whether different frequency bands of autonomous oscillations can be reproduced solely from differences in baseline activity.

\subsubsection{Phase coherence analysis}\label{sec:phase_coherence}

Because the population output $z(t)$ is the mean firing rate, the power of $z(t)$ in band $k$ depends not only on individual neuronal amplitudes but also on their phase relationships.
To quantify this effect, we define the instantaneous phase coherence in band $k$:
\begin{equation}
  R^{(k)}(t) = \left| \frac{1}{N} \sum_{i=1}^{N} A_i^{(k)}(t)\, e^{\mathrm{i}\phi_i^{(k)}(t)} \right|.
  \label{eq:phase_coherence}
\end{equation}
$R^{(k)}(t)$ corresponds to the instantaneous amplitude of the population signal in band $k$.
When the phases are perfectly aligned, $R^{(k)}(t) = \frac{1}{N}\sum_i A_i^{(k)}(t)$ (constructive interference); when they are randomly dispersed, $R^{(k)}(t) \approx 0$ (destructive interference).

The coherence ratio for band $k$ in mode $m$ is defined as
\begin{equation}
  \eta^{(m,k)} = \frac{\langle R^{(k)}(t) \rangle_{t \in \mathcal{T}_m}}
    {\left\langle \frac{1}{N}\sum_{i} A_i^{(k)}(t) \right\rangle_{t \in \mathcal{T}_m}}.
  \label{eq:coherence_ratio}
\end{equation}
A value $\eta \approx 1$ indicates constructive interference, while $\eta \approx 0$ indicates destructive interference.
This measure is used to quantitatively characterize rhythm switching driven by reorganization of phase relationships.

Furthermore, to directly test whether the amplitudes of individual neurons in a given band $k$ are preserved across modes, we compute the ratio of per-neuron band amplitudes between modes $m_1$ and $m_2$:
\begin{equation}
  \kappa_i^{(k; m_1, m_2)} = \frac{\bar{A}_i^{(m_2, k)}}{\bar{A}_i^{(m_1, k)}}.
  \label{eq:amplitude_ratio}
\end{equation}
$\kappa_i \approx 1$ indicates that the oscillation of neuron $i$ in band $k$ is maintained across modes, whereas $\kappa_i \approx 0$ indicates suppression.

\subsubsection{Causal silencing analysis}\label{sec:ablation}

To test causally whether high-frequency rhythms depend on short-time-constant neurons, we silenced subpopulations of neurons in each trained network and measured the effect on the band power of the population output.
Neurons were ranked by time constant and silenced progressively (in $10\%$ increments of the population) either from the longest time constant first (slow-first), from the shortest first (fast-first), or as random subsets of the corresponding size (control; averaged over five random subsets per fraction).
Two silencing methods were used.
In \emph{zero-ablation}, the output of a silenced neuron is set to zero, removing it from both the recurrent dynamics and the readout.
In \emph{freeze}, the output of a silenced neuron is instead clamped to its per-mode time-averaged firing rate $\bar{r}_i^{(m)}$; this removes the neuron's oscillatory contribution while preserving its mean (DC) contribution to the recurrent input and to the readout, thereby isolating its oscillatory role from its contribution to the baseline operating point (mechanism~B, Sect.~\ref{sec:baseline_results}).
For each silencing configuration we ran the network under the sequential four-mode protocol (as in Sect.~\ref{sec:analysis}, with $6000$-step intervals) and computed, for each mode, the in-band power of the population output relative to its value in the intact network.
This analysis was performed on all $n = 20$ trained networks.

\subsubsection{Connectivity structure and linearized dynamics}\label{sec:connectivity_methods}

To relate the learned recurrent weights $W_{\mathrm{rec}}$ and the linearized dynamics to the functional differentiation (Sect.~\ref{sec:connectivity_results}), we used three analyses.
(i)~\emph{Connectivity structure.} Neurons were partitioned into terciles by time constant, and we computed the mean absolute recurrent weight $|W_{\mathrm{rec}}|$ within the fast (short-$\tau$) tercile, within the slow (long-$\tau$) tercile, and between them (averaging the two off-diagonal directions), comparing within-fast to within-slow coupling across networks with a Wilcoxon signed-rank test. The result was unchanged when self-connections (diagonal entries) were excluded.
(ii)~\emph{Relaxation timescale.} For each network we located a fixed point $\bm{x}^*$ of the input-free dynamics by minimizing $\| f(\bm{x}) - \bm{x} \|^2$ (Eq.~\ref{eq:fp_objective}) starting from the origin $\bm{x} = \bm{0}$, formed the discrete-time Jacobian there (Eq.~\ref{eq:jacobian} with $r^*_j = \tanh(x^*_j)$), and defined the slowest relaxation time constant as $-1/\ln|\mu|$ for the stable eigenvalue of largest magnitude ($|\mu| < 1$), comparing it with the largest intrinsic time constant $\max_i \tau_i$ of the same network. The fixed point is unstable (it has oscillatory $|\mu| > 1$ modes; Sect.~\ref{sec:baseline_results}); this measure characterizes the slowest \emph{decaying} direction, not the oscillation frequency.
(iii)~\emph{Gain modulation.} For a representative run we evaluated the Jacobian at the mean activity $\bar{\bm{r}}^{(m)}$ (from a $4000$-step simulation) of the $\beta$ and $\gamma$ modes and linearly interpolated the gain profile $1 - \bar{r}_j^{2}$ between the two operating points (with $W_{\mathrm{rec}}$ and the leak rates held fixed), tracking the frequency $f = |\arg\mu|/(2\pi)\,F_s$ of the dominant unstable eigenvalue.

\subsubsection{Time-constant initialization control}\label{sec:reinit}

To test whether the functional differentiation depends on the initialization of the time constants (all set to $50$\,ms in the main experiments), we trained additional networks under two alternative schemes while keeping the time constants learnable: (i)~a per-neuron log-uniform initialization over $\tau \in [5, 300]$\,ms ($10$ networks), and (ii)~a uniform initialization at $\tau = 200$\,ms ($10$ networks). The architecture, task, and loss were otherwise identical. To limit the computational cost we used an early-stopping threshold of $\mathcal{L} < 10^{-5}$ (verified on the original networks to leave the amplitude--time-constant correlations essentially unchanged relative to $10^{-6}$, $|\Delta\rho| \le 0.06$). One log-uniform run failed to converge ($\mathcal{L}_{\min} = 1.2\times10^{-2}$) and was excluded, leaving $9$ and $10$ converged networks. For each network we recomputed the learned time-constant distribution and the per-band amplitude statistics and time-constant correlations as in Sect.~\ref{sec:amp_stats}--\ref{sec:tau_corr}, using the sequential four-mode protocol with $6000$-step intervals; the baseline statistics so recomputed reproduce those of Table~\ref{tab:amp_tau_stats}.

\section{Results}\label{sec:results}

\subsection{Training converges and reproduces target rhythms}\label{sec:training_success}

Of the $n = 20$ independent runs, 18 reached the early-termination criterion ($\mathcal{L} < 10^{-6}$) after $26{,}064 \pm 8{,}072$ iterations (range: 15{,}057--42{,}085).
The remaining two reached the 50{,}000-iteration limit but still achieved $\mathcal{L}_{\min} < 1.5 \times 10^{-6}$, indicating sufficient convergence (across all runs, $\bar{\mathcal{L}}_{\min} = 1.02 \times 10^{-6} \pm 1.27 \times 10^{-7}$).
All 20 runs are therefore used in the analyses that follow.

Figure~\ref{fig:task_overview} shows results from a representative run.
The population output $z(t)$ switches cleanly between rhythms in response to each pulse input (Fig.~\ref{fig:task_overview}b).
The power spectrum of each interval shows a sharp peak within the designated band (Fig.~\ref{fig:task_overview}c), confirming that the loss function concentrates power in the target band as intended.

\subsection{Amplitude distributions and their correlation with time constants}\label{sec:amp_tau}

\begin{figure*}[t]
  \centering
  \includegraphics[width=\textwidth]{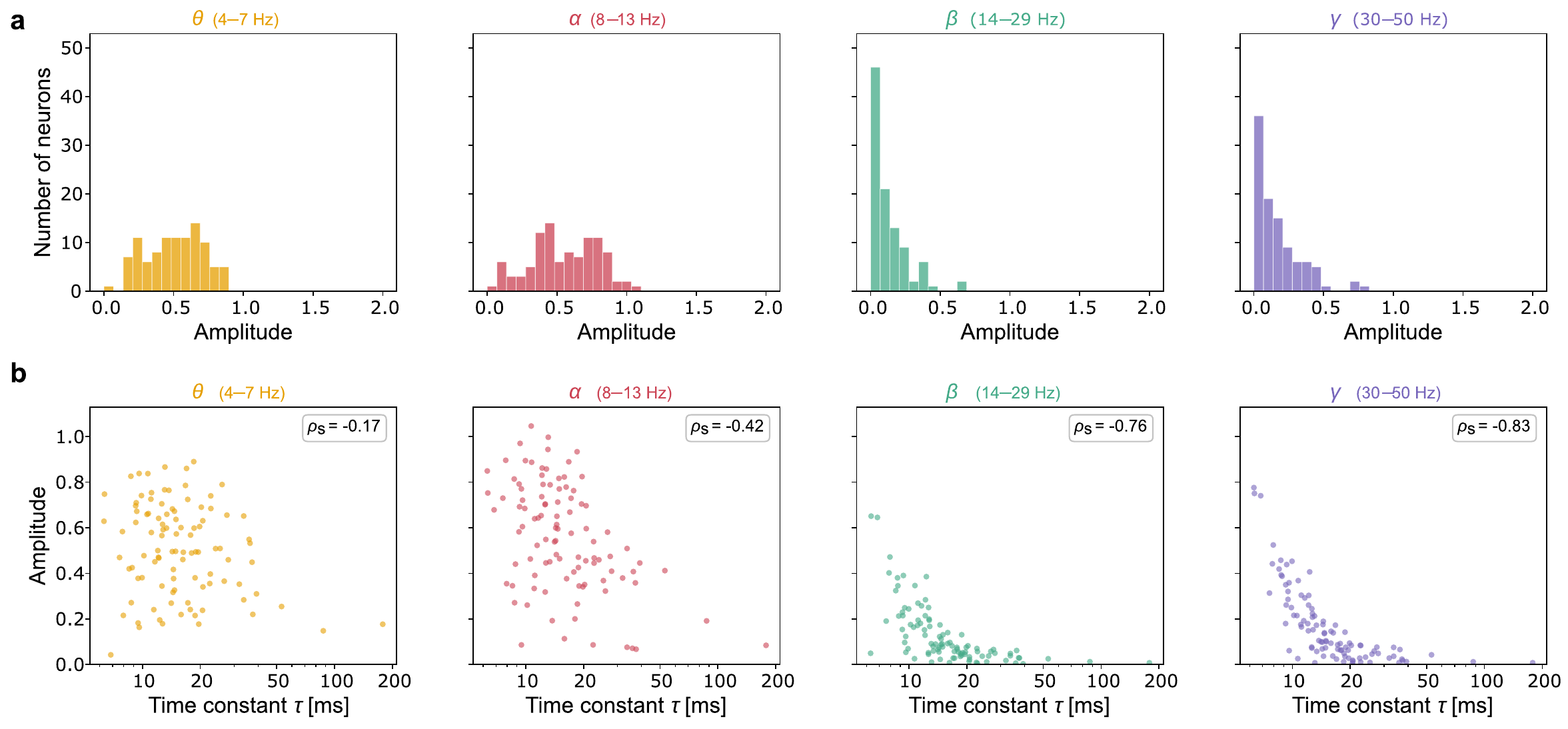}
  \caption{Frequency-dependent functional differentiation.
  (a) Distribution of matched-mode amplitudes for each mode (representative single run).
  (b) Scatter plot of time constant $\tau_i$ versus matched-mode amplitude (representative single run, four modes).}
  \label{fig:amp_tau}
\end{figure*}

We examined how the matched-mode amplitude $\bar{A}_i^{(m,m)}$ (Eq.~\ref{eq:band_amplitude}) is distributed across neurons in each mode.
Figure~\ref{fig:amp_tau}a shows the amplitude distribution from a representative run.
In the low-frequency modes ($\theta$, $\alpha$), amplitudes are distributed fairly uniformly and many neurons participate in the oscillation.
In contrast, the distributions for the high-frequency modes ($\beta$, $\gamma$) are concentrated near zero with heavy right tails, so that only a few neurons exhibit large amplitudes.

This trend is quantified by the statistics of the amplitude distribution (Table~\ref{tab:amp_tau_stats}).
Skewness measures the asymmetry of the distribution, with larger values indicating heavier right tails (i.e., the presence of a few large-amplitude neurons).
The Gini coefficient quantifies inequality, with $0$ denoting perfect equality and $1$ complete concentration.
Skewness increases monotonically with frequency, from $-0.30 \pm 0.20$ for $\theta$ to $1.60 \pm 0.61$ for $\gamma$, and the Gini coefficient follows the same trend, from $0.21 \pm 0.03$ ($\theta$) to $0.53 \pm 0.04$ ($\gamma$).
Higher-frequency rhythms are thus generated by a progressively smaller dominant subpopulation.

We next examined the relationship between the learned time constants $\tau_i=1/\lambda_i$ and the matched-mode amplitude (Fig.~\ref{fig:amp_tau}b).
Although the $\theta$ mode exhibits only a weak correlation between time constant and amplitude ($\rho_s^{\mathrm{amp}} = -0.15 \pm 0.11$), the negative correlation strengthens with mode frequency, reaching $\rho_s^{\mathrm{amp}} = -0.82 \pm 0.06$ in $\gamma$.
That is, whereas high-frequency rhythms selectively recruit neurons with short time constants, low-frequency rhythms engage the population broadly, independently of time constant.

\begin{table*}[t]
  \centering
  \caption{Amplitude distribution statistics and correlations with time constant (mean $\pm$ SD over $n = 20$ runs).}
  \label{tab:amp_tau_stats} 
  \begin{tabular}{lccccc}
    \toprule
    Mode & Skewness & Gini & $\bar{d}$ & $\rho_s^{\mathrm{amp}}(\tau, \bar{A})$ & $\rho_s^{\mathrm{sync}}(\tau, d)$ \\
    \midrule
    $\theta$ & $-0.30 \pm 0.20$ & $0.21 \pm 0.03$ & $36.1 \pm 9.6$  & $-0.15 \pm 0.11$ & $-0.15 \pm 0.11$ \\
    $\alpha$ & $0.32 \pm 0.83$  & $0.30 \pm 0.08$ & $26.3 \pm 17.9$ & $-0.43 \pm 0.10$ & $-0.43 \pm 0.10$ \\
    $\beta$  & $1.13 \pm 0.38$  & $0.42 \pm 0.05$ & $2.4 \pm 2.9$   & $-0.68 \pm 0.11$ & $-0.67 \pm 0.11$ \\
    $\gamma$ & $1.60 \pm 0.61$  & $0.53 \pm 0.04$ & $2.1 \pm 1.3$   & $-0.82 \pm 0.06$ & $-0.81 \pm 0.07$ \\
    \bottomrule
  \end{tabular}
\end{table*}

\subsection{Frequency-dependent synchronization structure}\label{sec:sync_results}

\begin{figure*}[t]
  \centering
  \includegraphics[width=\textwidth]{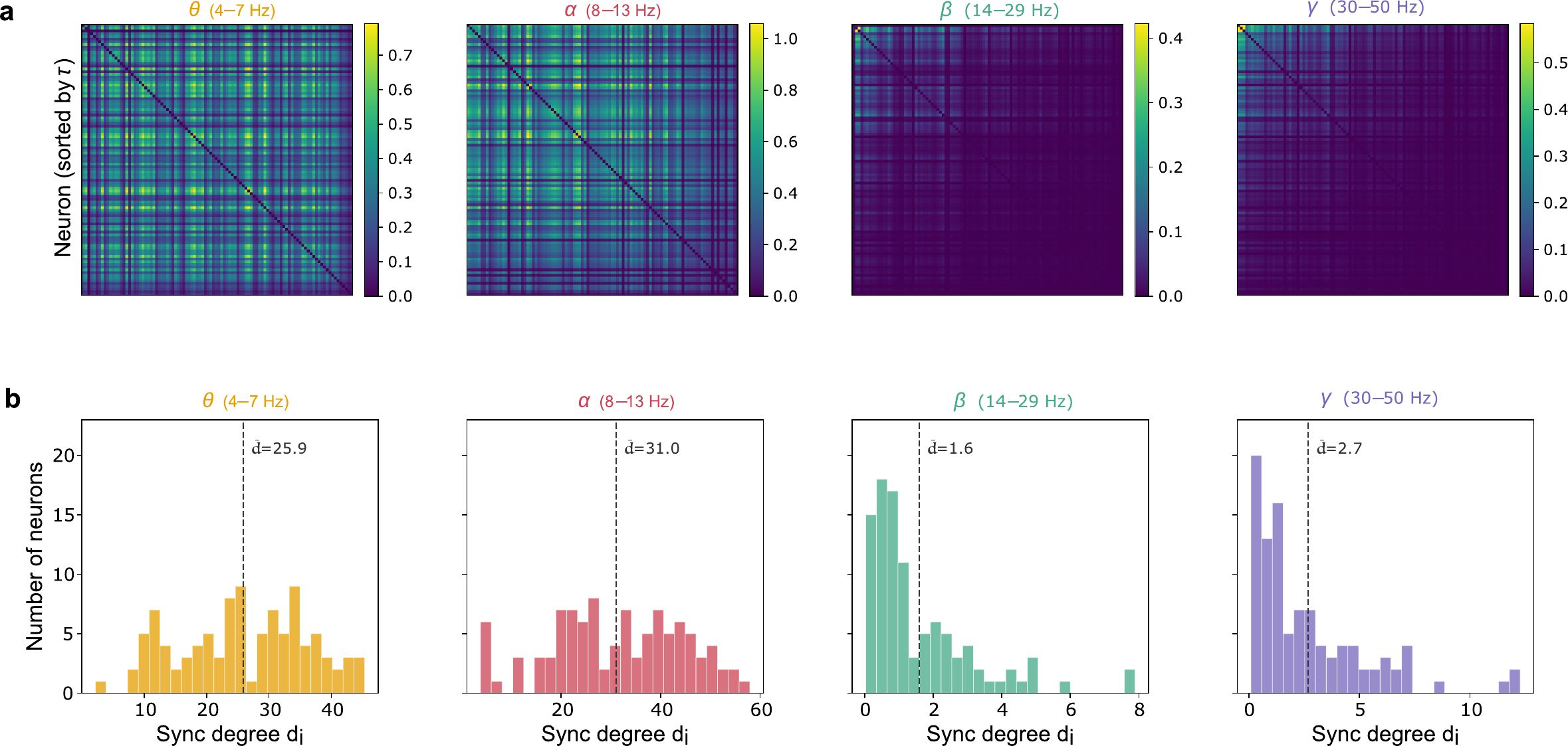}
  \caption{Frequency dependence of the synchronization structure.
  (a) Absolute sync strength matrix $S_{ij}^{(m)}$ (representative single run; neurons are sorted in ascending order of $\tau_i$, and the same ordering is used for all four modes. The color scale is set independently for each mode).
  (b) Distribution of the sync degree $d_i$ in each mode (representative single run).}
  \label{fig:sync}
\end{figure*}

We computed the absolute sync strength $S_{ij}^{(m)}$ (Eq.~\ref{eq:abs_sync}) for each mode and visualized it with the neurons sorted in ascending order of time constant $\tau_i$ (Fig.~\ref{fig:sync}a).
In the $\theta$ and $\alpha$ modes, high sync strength is broadly distributed across neuron pairs, spanning the entire matrix.
In the $\beta$ and $\gamma$ modes, by contrast, high values appear only in the upper-left region (pairs of neurons with short time constants), indicating that the synchronized population comprises a small number of fast neurons.

The distribution of the sync degree $d_i$ (Eq.~\ref{eq:sync_degree}; Fig.~\ref{fig:sync}b) shows the same frequency dependence as the matched-mode amplitude distribution (Fig.~\ref{fig:amp_tau}a): broad for low-frequency modes and concentrated near zero with a heavy right tail for high-frequency modes.
The mean sync degree drops by over an order of magnitude from $36.1 \pm 9.6$ in the $\theta$ mode to $2.1 \pm 1.3$ in the $\gamma$ mode (Table~\ref{tab:amp_tau_stats}). 
The correlation between sync degree and time constant, $\rho_s^{\mathrm{sync}}$, strengthens toward higher-frequency modes, paralleling the amplitude correlations (Table~\ref{tab:amp_tau_stats}).

Because $S_{ij}$ is amplitude-weighted (Eq.~\ref{eq:abs_sync}), the sync degree $d_i$ is almost perfectly rank-correlated with the matched-mode amplitude within each network (Spearman $\rho \ge 0.99$ in all four modes). Consequently $\rho_s^{\mathrm{sync}}$ and $\rho_s^{\mathrm{amp}}$ are nearly identical (Table~\ref{tab:amp_tau_stats}): at the single-neuron level the sync degree carries essentially the same information as the amplitude, and its distinct contribution lies instead in the pairwise organization of $S_{ij}$ (Fig.~\ref{fig:sync}a), which the row sum collapses.

\subsection{Inter-mode population similarity and run-to-run variability}\label{sec:intermode_results}

\begin{figure*}[t]
  \centering
  \includegraphics[width=\textwidth]{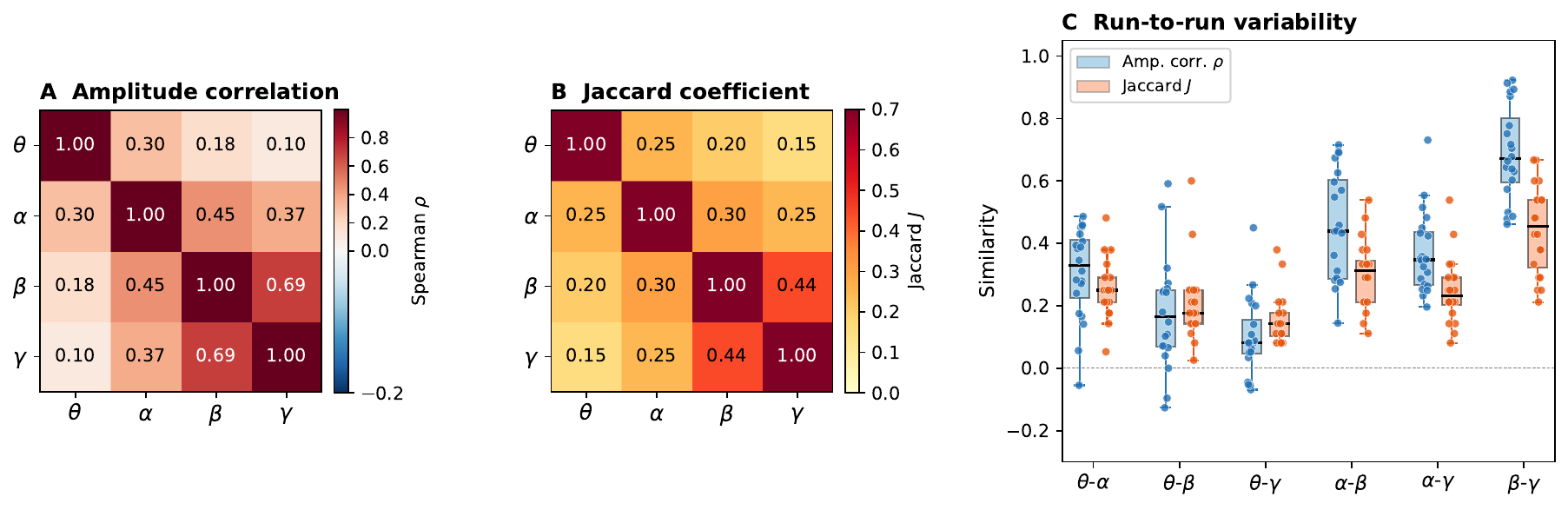}
  \caption{Inter-mode population similarity.
  (a) Spearman correlation matrix of matched-mode amplitudes (mean over $n = 20$ runs).
  (b) Jaccard coefficient matrix for the top 20\% of neurons by sync degree (mean over $n = 20$ runs).
  (c) Run-to-run distribution of amplitude correlations and Jaccard coefficients for all six mode pairs.}
  \label{fig:intermode}
\end{figure*}

We next asked how much the neuronal populations carrying each oscillation overlap across modes.
Figure~\ref{fig:intermode}a shows the Spearman correlation matrix of the matched-mode amplitude vectors, $\rho_{m_1 m_2}^{\mathrm{amp}}$ (Eq.~\ref{eq:amp_corr}), averaged across runs.

The $\theta$-mode amplitude pattern correlates weakly with every other mode ($\rho_{\theta,\cdot}^{\mathrm{amp}} = 0.10$--$0.30$), and the relative independence of the $\theta$-rhythm population is largely preserved across the $n = 20$ runs, although a few runs exhibit a moderately elevated $\theta$--$\beta$ correlation (Fig.~\ref{fig:intermode}c).
The $\alpha$--$\beta$ and $\beta$--$\gamma$ pairs, in contrast, show moderate positive correlations ($\rho_{\alpha,\beta}^{\mathrm{amp}} = 0.45 \pm 0.17$, $\rho_{\beta,\gamma}^{\mathrm{amp}} = 0.69 \pm 0.15$), indicating partial sharing of the neuronal populations among the high-frequency modes.

The Jaccard coefficient between the top 20\% of neurons by sync degree (Fig.~\ref{fig:intermode}b) shows the same structure in which $J$ is low ($0.15$--$0.25$) between $\theta$ and the other modes but relatively high between $\beta$ and $\gamma$ ($0.44 \pm 0.14$).

Notably, the similarity patterns among $\alpha$, $\beta$, and $\gamma$ vary substantially across training runs (Fig.~\ref{fig:intermode}c). For example, the amplitude correlation between $\beta$ and $\gamma$ ranges from $0.46$ to $0.92$ (SD $= 0.15$).
While $\theta$ remains largely independent across runs, the sharing structure among $\alpha$, $\beta$, and $\gamma$ depends on training initialization, giving rise to a diverse set of solutions.
This run-to-run variability is partly organized by the learned time constants: across the $20$ networks, the $\beta$--$\gamma$ amplitude-pattern correlation scales with the median learned time constant (Spearman $\rho = 0.45$, $p = 0.048$; Appendix~\ref{secC1}, Fig.~\ref{fig:outlier_tau}). Networks that learn a lower median time constant possess a larger pool of fast neurons and can allocate distinct subpopulations to $\beta$ and $\gamma$ (lower sharing; e.g.\ the lowest-sharing networks, with $\beta$--$\gamma$ correlation near $0.5$), whereas networks with a higher median time constant share the smaller fast pool across both bands (higher sharing). A similar but weaker and non-significant tendency is seen for the $\theta$--$\beta$ pair ($\rho = 0.26$, $p = 0.27$). 

\subsection{Baseline shifts drive rhythm switching}\label{sec:baseline_results}

\begin{figure*}[t]
  \centering
  \includegraphics[width=\textwidth]{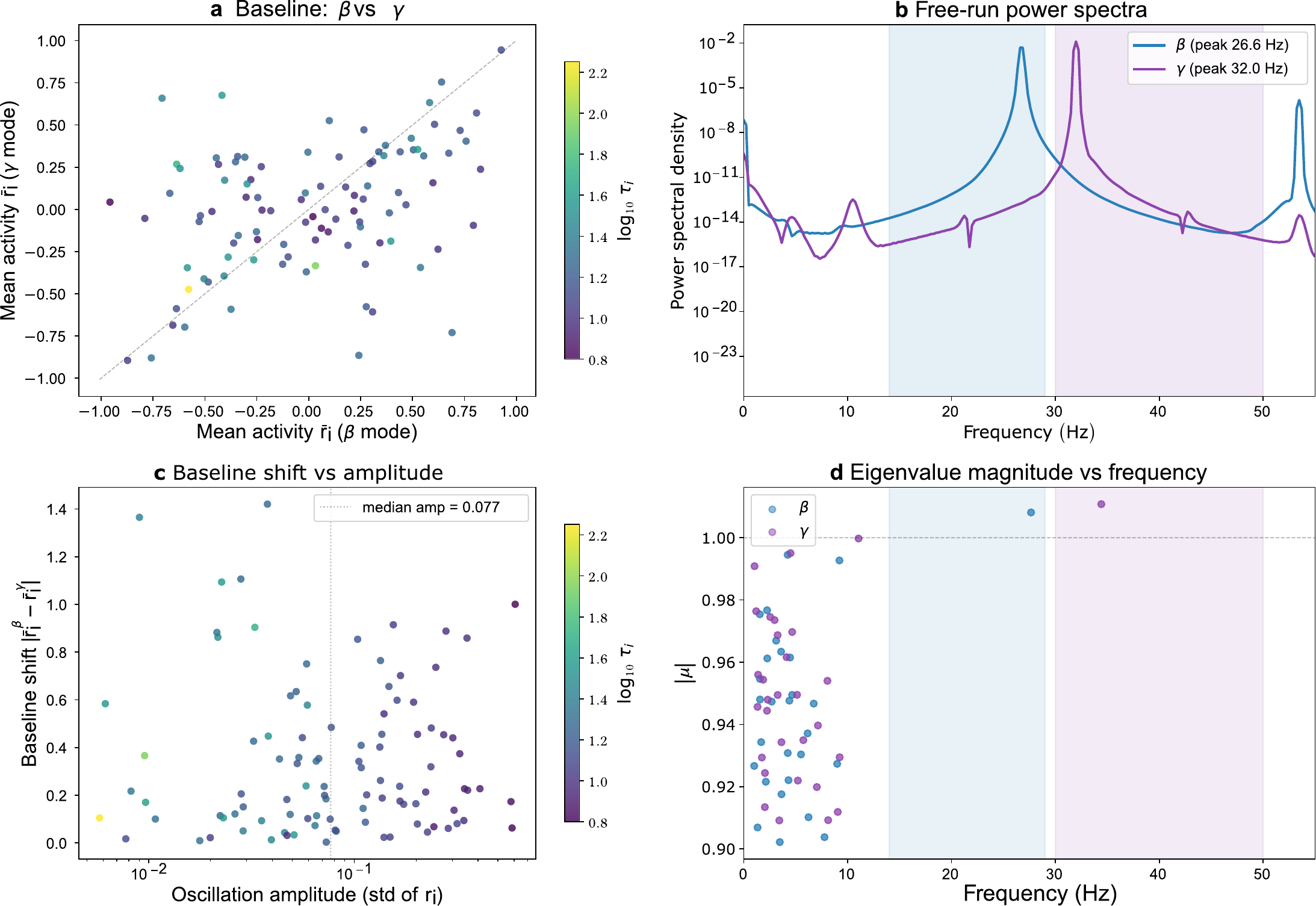}
  \caption{Baseline shift and fixed-point analysis ($\beta / \gamma$, representative single run).
  (a) Scatter plot of each neuron's time-averaged activity $\bar{r}_i^{(\beta)}$ versus $\bar{r}_i^{(\gamma)}$ during the $\beta$ and $\gamma$ modes.
  Color encodes the time constant $\tau_i$. Deviation from the diagonal corresponds to the magnitude of the baseline shift.
  (b) Power spectrum of the population output in a free-run simulation initialized from the mean activity of each mode.
  The $\beta$ initial state peaks at 26.6\,Hz and the $\gamma$ initial state peaks at 32.0\,Hz.
  Shaded regions indicate the target rhythm bands (blue: $\beta$, 14--29\,Hz; purple: $\gamma$, 30--50\,Hz).
  (c) Relationship between baseline shift $|\bar{r}_i^\beta - \bar{r}_i^\gamma|$ and oscillation amplitude.
  The vertical dashed line marks the median amplitude. The mean shifts for the two groups are nearly equal, indicating that the shift extends across the entire network.
  (d) Eigenvalues of the discrete-time Jacobian. The horizontal axis is the oscillation frequency estimated from the eigenvalue, and the vertical axis is the eigenvalue magnitude $|\mu|$.
  Eigenvalues with $|\mu| > 1$ (dashed line) indicate unstable oscillatory modes, located at 27.7\,Hz for $\beta$ and 34.4\,Hz for $\gamma$.}
  \label{fig:baseline}
\end{figure*}

We next asked how the network changes the oscillation frequency when the neuronal populations carrying the oscillation do not undergo substantial turnover between modes, as is typical for the $\beta$--$\gamma$ pair.
Inspection of individual neuronal outputs reveals that, alongside changes in amplitude and frequency, the mean activity level (baseline) also shifts between modes.
The shift is pronounced even for slow, low-amplitude neurons.
To probe the dynamical significance of this baseline shift, we perform a fixed-point analysis. Because the fixed-point and eigenvalue analyses are carried out per run and do not summarize compactly across runs, we focus on a single exemplar. 
We select a run in which the $\beta$ and $\gamma$ populations overlap substantially ($J(\beta,\gamma) = 0.667$), a critical test case where population turnover cannot account for the frequency change.
Figure~\ref{fig:baseline}a shows a scatter plot of the time-averaged activity $\bar{r}_i^{(\beta)}$ versus $\bar{r}_i^{(\gamma)}$ of each neuron in this run. Many neurons deviate substantially from the diagonal, confirming that the $\beta$--$\gamma$ baseline shift extends across the entire network.

\paragraph{Existence and stability of fixed points}
Starting the L-BFGS-B optimization from initial conditions near the time-averaged activity of each mode, the search yielded, in each mode, a fixed point satisfying $q < 10^{-18}$.
Searches from six perturbed initial conditions near the mean activity of each mode converged to a single solution, suggesting that the fixed point is locally isolated in the neighborhood of the baseline.

Eigenvalue analysis of the Jacobian at each fixed point confirms that both are unstable (Fig.~\ref{fig:baseline}d).
At the $\beta$ fixed point, the only oscillatory instability (a complex-conjugate eigenvalue pair with $|\mu| > 1$) lies at $27.7$\,Hz (within the $\beta$ band) with $|\mu| = 1.008$, and similarly, the $\gamma$ fixed point has a single unstable oscillatory pair at $34.4$\,Hz (within the $\gamma$ band) with $|\mu| = 1.011$.
Each fixed point is thus destabilized only at a frequency within its corresponding band, showing that the baseline shift switches the dominant unstable oscillatory mode.

Free-run simulations initialized at the mean activity of each mode reproduce autonomous oscillations in distinct frequency bands (Fig.~\ref{fig:baseline}b).
The resulting power spectra lie within the corresponding rhythm bands and agree well with the oscillation frequencies predicted by the Jacobian eigenvalues.
This suggests that the linear instability near the fixed point approximately sets the frequency of the limit cycle.

\paragraph{Structure of the baseline shift}
Figure~\ref{fig:baseline}c plots the baseline shift $|\bar{r}_i^{\beta} - \bar{r}_i^{\gamma}|$ of each neuron against its oscillation amplitude (the standard deviation of $r_i(t)$).
The baseline shift is a collective phenomenon involving the entire network, including not only the large-amplitude neurons but also low-amplitude neurons, and this network-wide shift of the operating point selects the unstable oscillatory mode.
This corresponds to mechanism~B described in the Introduction.

\subsection{Fast neurons are the causal substrate of high-frequency rhythms}\label{sec:ablation_results}

The analyses above establish a correlational link between short time constants and participation in high-frequency rhythms. To test this link causally, we silenced neurons ranked by time constant and measured the retained in-band power of the population output (Sect.~\ref{sec:ablation}; Fig.~\ref{fig:ablation}).

Using the freeze method, which removes a neuron's oscillation while preserving its baseline contribution, silencing the fastest neurons abolished the high-frequency rhythms almost immediately: freezing only the fastest $10\%$ of neurons reduced both $\beta$ and $\gamma$ power to essentially zero (median retained power $0.00$ for both, across the $n=20$ networks).
In contrast, freezing the same fraction of the \emph{slowest} neurons left the high-frequency rhythms largely intact ($\gamma$: median retained power $0.96$; $\beta$: $0.79$), and $\gamma$ power remained substantial even after freezing the slowest $40\%$ of the population (median $0.50$).
This asymmetry weakened toward lower frequencies ($\alpha$: $0.40$ for slow-first vs.\ $0.00$ for fast-first at $10\%$) and vanished for $\theta$, where silencing any $10\%$ of neurons---fast, slow, or random---abolished or strongly suppressed the rhythm, consistent with the distributed, non-time-constant-specific generation of $\theta$ (Sect.~\ref{sec:amp_tau}).

The freeze method preserves each mode's baseline while removing a neuron's oscillation, and therefore isolates the oscillatory contribution: it identifies the fast subpopulation as the primary oscillatory substrate of $\beta$ and $\gamma$, in agreement with the amplitude, synchronization, and connectivity results. 
However, there are two points to note.
First, because the frozen neurons are clamped to their mode-specific baseline, this shows slow-neuron oscillations to be unnecessary given that baseline; fully removing the slow neurons instead also reduces $\beta$/$\gamma$ power (Fig.~\ref{fig:ablation_supp}), because it additionally disrupts the baseline that positions the operating point (mechanism~B, Sect.~\ref{sec:baseline_results}). 
Second, because the silencing is applied from the onset of each interval, including the transition into the mode, the collapse of $\beta$ and $\gamma$ under fast-neuron silencing may reflect a role of the fast neurons in establishing these rhythms, not only in maintaining them.

\subsection{Phase interference controls band power}\label{sec:phase_results}

The analyses in Sects.~\ref{sec:amp_tau}--\ref{sec:baseline_results} focused on changes in single-neuron amplitude or activity level.
Because the population output $z(t) = \frac{1}{N}\sum_i r_i(t)$ is the population mean, however, band-specific power can also be shaped by inter-neuronal phase relationships.
Here we examine the role of phase interference using the coherence ratio $\eta^{(m,k)}$ (Eq.~\ref{eq:coherence_ratio}).

We computed the coherence ratio for all bands and modes across the $n = 20$ runs.
The diagonal components ($k = m$) were low for low-frequency modes and high for high-frequency modes
($\eta^{(\theta,\theta)} = 0.10 \pm 0.03$,
$\eta^{(\alpha,\alpha)} = 0.14 \pm 0.08$,
$\eta^{(\beta,\beta)} = 0.41 \pm 0.12$,
$\eta^{(\gamma,\gamma)} = 0.52 \pm 0.09$; mean $\pm$ SD).
That is, low-frequency rhythms are produced by many neurons contributing at diverse phases, whereas high-frequency rhythms are produced by a small number of neurons operating in a coherent phase, consistent with the amplitude-concentration result of Sect.~\ref{sec:amp_tau}.

\begin{figure*}[t]
  \centering
  \includegraphics[width=\textwidth]{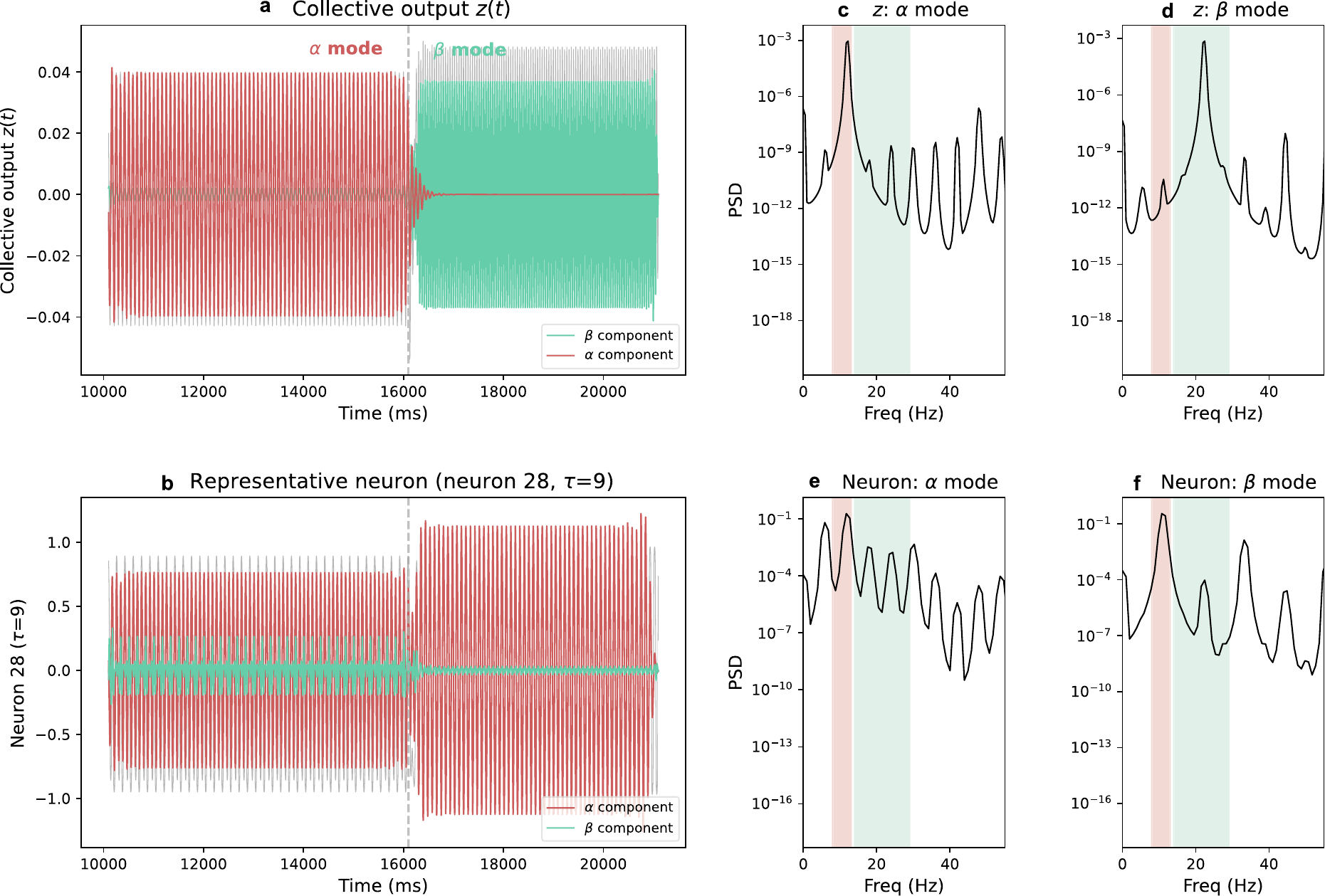}
  \caption{Visualization of the phase-interference mechanism ($\alpha \to \beta$ transition, representative phase-interference-type run).
  (a) Continuous time series of the population output $z(t)$.
  Red and green traces denote the $\alpha$- and $\beta$-band bandpass-filtered components, respectively.
  After the transition, the $\alpha$ component disappears and the $\beta$ component emerges.
  (b) Activity of a representative neuron (neuron 28, $\tau = 9$).
  The $\alpha$ component (red) is maintained---indeed amplified---in the $\beta$ mode.
  (c,d) Power spectra of the population output in each mode.
  The $\alpha$ peak present in the $\alpha$ mode (c) disappears in the $\beta$ mode (d).
  (e,f) Power spectra of the representative neuron in each mode.
  The $\alpha$-band power is nearly identical across the two modes (power ratio $= 2.03$), showing that the $\alpha$ oscillation is not suppressed at the single-neuron level.}
  \label{fig:phase}
\end{figure*}

To assess whether the $\alpha$-band amplitude of individual neurons actually changes during the $\alpha \to \beta$ transition, we next examined the amplitude ratio $\kappa_i^{(\alpha;\,\alpha,\beta)}$.
The distribution of the neuron-averaged $\bar{\kappa}$ across runs was bimodal (Fig.~\ref{fig:sup_kappa}a,b).
Using $\bar{\kappa} > 0.1$ as a threshold, 13 runs were classified as the phase-interference type and 7 as the amplitude-suppression type.
The amplitude-suppression type implements rhythm switching through mechanisms~A and~B, whereas the phase-interference type exhibits qualitatively different behavior.
Figure~\ref{fig:phase} shows a representative example.
During the $\alpha \to \beta$ transition, the $\alpha$ component disappears from the population output $z(t)$ and a $\beta$ component emerges (Fig.~\ref{fig:phase}a,c,d).
Yet at the single-neuron level, the $\alpha$-band oscillation actually \emph{increases} in the $\beta$ mode (representative neuron $\alpha$-band power ratio $= 2.03$; Fig.~\ref{fig:phase}b,e,f).
In this run, 98 of the 100 neurons had $\kappa_i > 0.5$, with a neuron-averaged value of $\bar{\kappa} = 1.96 \pm 0.99$.
Despite this, the coherence ratio of the $\alpha$ band in the $\beta$ mode was extremely low ($\eta^{(\beta,\alpha)} = 0.0002$), showing that destructive phase interference cancels the $\alpha$ component from the population signal.
This corresponds to mechanism~C described in the Introduction.

Thus, for the same task and loss function, rhythm switching can be implemented internally by either amplitude-based or phase-interference-based solutions.
Because the loss function (Eq.~\ref{eq:loss}) evaluates only the power spectrum of the population output, both single-neuron amplitude changes and inter-neuronal phase reorganization are available as degrees of freedom, and training initialization determines which solution is selected.
The microscopic mechanisms that achieve a macroscopic objective are therefore not uniquely determined, consistent with the diversity of learned solutions reported in Sect.~\ref{sec:intermode_results}.

\subsection{Structural and dynamical basis: recurrent connectivity and gain modulation}\label{sec:connectivity_results}

\begin{figure*}[t]
  \centering
  \includegraphics[width=\textwidth]{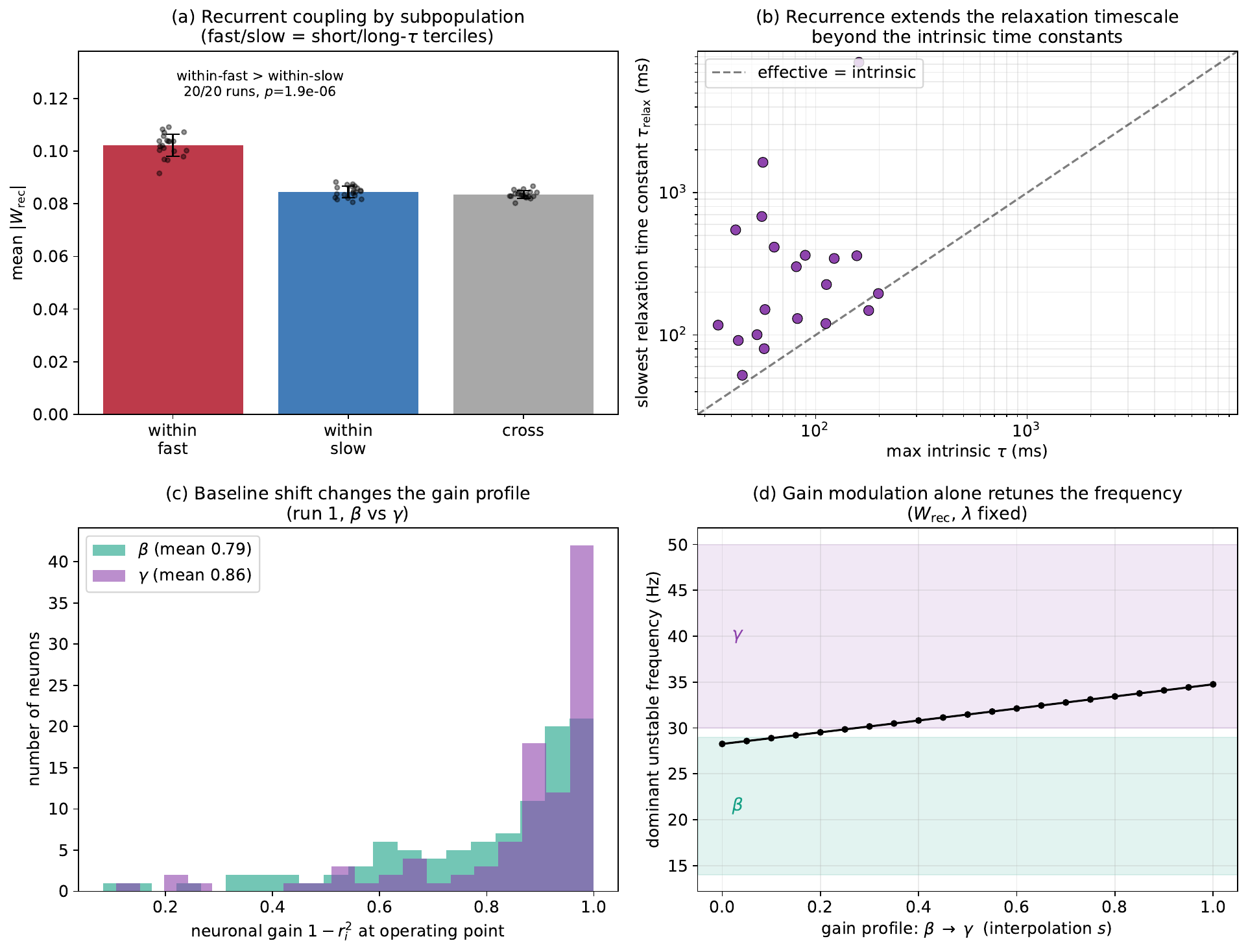}
  \caption{Structural and dynamical basis of the learned rhythms.
  (a) Mean absolute recurrent weight $|W_{\mathrm{rec}}|$ within the fast (short-$\tau$) subpopulation, within the slow (long-$\tau$) subpopulation, and between them (terciles of the time-constant distribution; bars and error bars are mean $\pm$ SD over $n=20$ networks, dots are individual networks). Within-fast coupling is the largest in all $20$ networks.
  (b) Slowest relaxation (decaying, $|\mu|<1$) time constant of the input-free dynamics linearized about a fixed point, versus the longest intrinsic time constant, for each network (log--log; dashed line is equality). Most networks ($18/20$) lie above the diagonal: recurrence extends the relaxation timescale beyond the intrinsic time constants.
  (c) Distribution of the neuronal gain $1 - \bar{r}_i^{2}$ at the $\beta$ and $\gamma$ operating points (representative run~1). The baseline shift between modes changes the gain profile.
  (d) Dominant unstable eigenfrequency as the gain profile is interpolated from the $\beta$ to the $\gamma$ operating point, with $W_{\mathrm{rec}}$ and the leak rates held fixed. Gain modulation alone moves the frequency from the $\beta$ band (green) to the $\gamma$ band (purple).}
  \label{fig:connectivity}
\end{figure*}

The mechanisms characterized so far were described at the level of neuronal activity. To probe their structural and dynamical foundations---and because the effective time constant of a recurrent circuit is shaped by its connectivity rather than by single-neuron properties alone \citep{marti2018, stern2023, mazzucato2022}---we analyzed the learned recurrent weight matrix $W_{\mathrm{rec}}$ and the linearized (Jacobian) dynamics of the trained networks (Fig.~\ref{fig:connectivity}).

First, the recurrent connectivity is organized with respect to the time-constant-based functional differentiation (Fig.~\ref{fig:connectivity}a). Partitioning neurons into terciles by time constant, the mean absolute recurrent weight within the fast (short-$\tau$) subpopulation ($0.102 \pm 0.004$) substantially exceeds that within the slow subpopulation ($0.085 \pm 0.002$) and that between the two ($0.084 \pm 0.001$); within-fast coupling was the largest in all $20$ networks (Wilcoxon signed-rank $p = 1.9\times10^{-6}$). The fast neurons that carry the high-frequency rhythms thus form a more strongly interconnected recurrent module, providing a structural substrate for mechanism~A.

Second, the recurrent coupling generates slow collective \emph{relaxation} timescales that in most networks exceed any intrinsic neuronal time constant (Fig.~\ref{fig:connectivity}b). Linearizing the input-free dynamics about a fixed point (located from the origin; Sect.~\ref{sec:connectivity_methods}), the slowest \emph{decaying} (stable, $|\mu|<1$) eigenmode has a relaxation time constant (median $211$\,ms across networks) that exceeds the longest intrinsic time constant of the same network (median $72$\,ms) in $18$ of $20$ networks, by a median factor of $2.5$. The fixed point is itself unstable: its oscillatory ($|\mu|>1$) eigenmodes are the rhythm generators analyzed in Sect.~\ref{sec:baseline_results} (Fig.~\ref{fig:baseline}), whereas here we characterize the complementary, slowest relaxation timescale. This relaxation timescale is therefore a collective property shaped by the recurrent connectivity, consistent with the theoretical expectation that single-neuron time-constant diversity alone does not determine network timescales \citep{stern2023, marti2018, mazzucato2022}, and it provides a substrate for the low-frequency dynamics generated by recurrence rather than by intrinsically slow neurons.

Third, the baseline-shift mechanism (mechanism~B) operates through gain modulation of this fixed connectivity (Fig.~\ref{fig:connectivity}c,d). Linearizing about the mean activity $\bar{\bm{r}}^{(m)}$ of each mode (Eq.~\ref{eq:jacobian} evaluated at $\bar{\bm{r}}^{(m)}$ rather than at an exact fixed point), the effective coupling from neuron $j$ to neuron $i$ is $\lambda_i\,(1 - \bar{r}_j^{2})\,(W_{\mathrm{rec}})_{ij}$, so the neuronal gain $1 - \bar{r}_j^{2}$ scales the structural weight at the operating point. Between the $\beta$ and $\gamma$ modes the network-wide baseline shift changes this gain profile (mean gain $0.79$ vs.\ $0.86$ in the representative run; Fig.~\ref{fig:connectivity}c). Holding $W_{\mathrm{rec}}$ and the leak rates fixed and interpolating the gain profile from the $\beta$ to the $\gamma$ operating point moves the dominant unstable eigenfrequency continuously from the $\beta$ band into the $\gamma$ band (Fig.~\ref{fig:connectivity}d). The frequency retuning attributed to the baseline shift is thus mediated by gain modulation of the recurrent interactions, the same principle by which local connectivity and gain reshape effective timescales in cortical circuit models \citep{marti2018, mazzucato2022}.

\subsection{Robustness to time-constant initialization}\label{sec:reinit_results}

Because the results above emerged from networks initialized with a uniform time constant of $50$\,ms, we asked whether they depend on this choice by training additional networks with per-neuron log-uniform ($5$--$300$\,ms) or uniform $200$\,ms initializations, keeping the time constants learnable (Sect.~\ref{sec:reinit}). The frequency-dependent differentiation is robust: in all schemes the amplitude--time-constant correlation strengthens monotonically from $\theta$ to $\gamma$, and the networks drive the bulk of the time constants to short values regardless of the initialization (median learned $\tau = 13.5$, $13.9$, and $18.3$\,ms for the $50$\,ms, log-uniform, and $200$\,ms schemes), so the concentration of short time constants is not an artifact of the $50$\,ms initialization. Full results---including how the low-frequency correlations depend on the availability of slow neurons---are given in Appendix~\ref{secD1} (Figs.~\ref{fig:reinit_metrics}, \ref{fig:reinit_tau}).

\section{Discussion}\label{sec:discussion}

\subsection{Summary and a three-mechanism framework}\label{sec:summary_mechanisms}

We trained a leaky integrator RNN with adaptive time constants on a four-band rhythm-switching task and systematically analyzed the functional differentiation that emerged inside the network.
The main findings can be summarized as follows.
(i) Low-frequency rhythms ($\theta$, $\alpha$) are produced by a distributed mechanism in which many neurons contribute with amplitude distributed relatively broadly, whereas high-frequency rhythms ($\beta$, $\gamma$) are produced by a concentrated mechanism in which amplitude is dominated by a small dominant subpopulation (Sect.~\ref{sec:amp_tau}).
(ii) This functional differentiation corresponds systematically to the learned time constants: neurons with shorter time constants are selectively recruited for higher-frequency bands (Sects.~\ref{sec:amp_tau}, \ref{sec:sync_results}).
(iii) For the same task, the relative independence of the $\theta$-rhythm population is largely preserved across runs, while the sharing patterns among $\alpha$, $\beta$, and $\gamma$ vary considerably across training runs (Sect.~\ref{sec:intermode_results}).

\begin{figure*}[t]
  \centering
  \includegraphics[width=\textwidth]{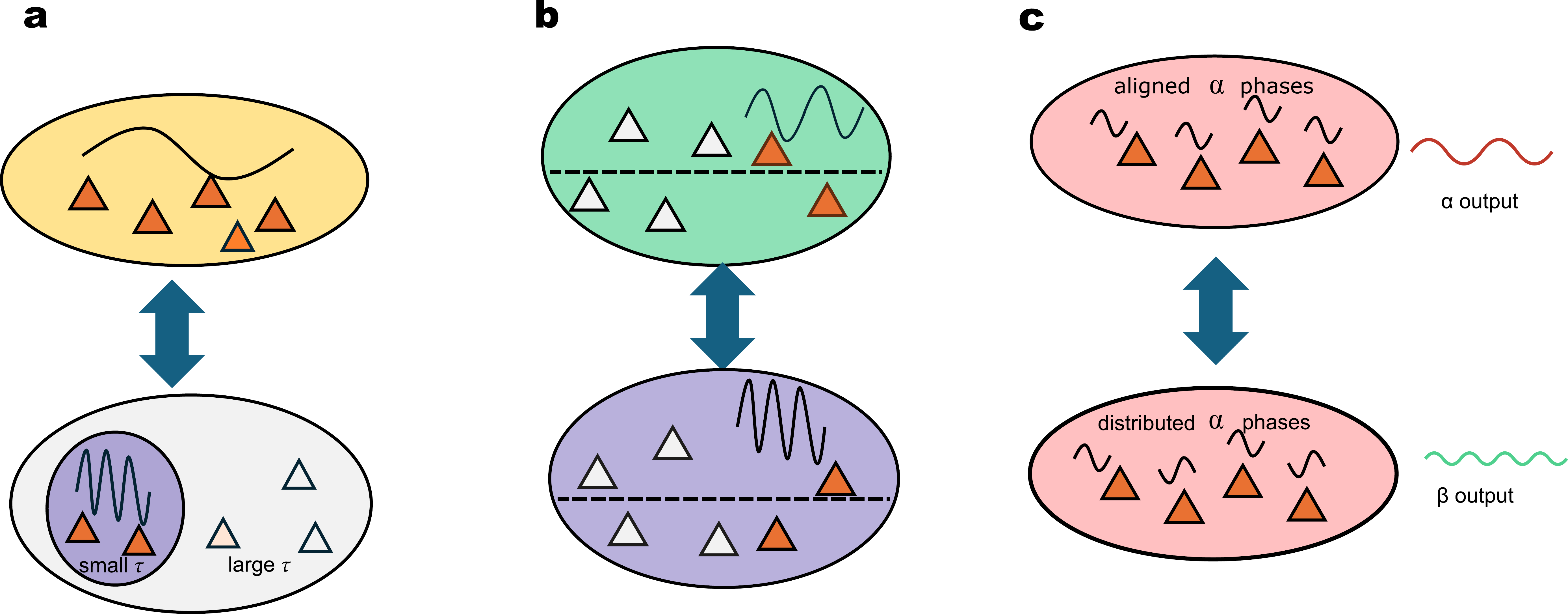}
  \caption{Schematic of the three rhythm-switching mechanisms (see Sect.~\ref{sec:summary_mechanisms} for details).
  Triangles represent neurons; filled (orange) triangles denote large oscillation amplitude and open triangles small amplitude.
  (a) Population turnover ($\theta \leftrightarrow \gamma$): the oscillating subpopulation differs between modes.
  (b) Baseline shift ($\beta \leftrightarrow \gamma$): the oscillating population is largely shared, but the mean activity of each neuron (vertical distance from the dashed line) shifts between modes.
  (c) Phase interference ($\alpha \leftrightarrow \beta$): individual neurons maintain their $\alpha$-band oscillation, but inter-neuronal phase alignment differs between modes.}
  \label{fig:mechanisms}
\end{figure*}

These findings suggest that rhythm switching is not governed by a single principle: at least three distinct mechanisms coexist and operate in combination (Fig.~\ref{fig:mechanisms}).

\textbf{Mechanism~A: population turnover} (Sect.~\ref{sec:amp_tau}).
In the $\theta$ rhythm, many neurons participate in the oscillation, whereas in the $\gamma$ rhythm a small group of short-time-constant neurons forms the synchronized population and most neurons exhibit only weak activity.
The size and composition of the oscillating population therefore differ qualitatively between low and high frequencies, so that mode transitions are accompanied by turnover of the active population.
Silencing experiments confirm that this dependence is causal: freezing the fastest subpopulation abolishes the $\beta$ and $\gamma$ rhythms, whereas freezing an equal fraction of the slowest neurons leaves them largely intact (Sect.~\ref{sec:ablation_results}).

\textbf{Mechanism~B: baseline shift} (Sect.~\ref{sec:baseline_results}).
This phenomenon is typical of $\beta$--$\gamma$ transitions.
The set of neurons with large amplitude does not change substantially between the two modes, but the network-wide mean activity level shifts, repositioning the operating point of the dynamical system near a different unstable fixed point in phase space and thereby altering the oscillation frequency.
Our fixed-point analysis showed that the fixed points near the mean activity of the $\beta$ and $\gamma$ modes are destabilized at frequencies within the corresponding bands.
The magnitude of the baseline shift is approximately uniform across the network regardless of oscillation amplitude, so the frequency is set by a collective repositioning of the operating point rather than by the high-amplitude neurons alone.

\textbf{Mechanism~C: phase interference} (Sect.~\ref{sec:phase_results}).
This phenomenon appears in a subset of runs during $\alpha \to \beta$ transitions.
The $\alpha$-band oscillation of individual neurons is maintained even in the $\beta$ mode, but their phases disperse so that destructive interference cancels the $\alpha$ component in the population signal, while the higher-frequency $\beta$ component appears in the output.
This mechanism exploits the fact that the loss function evaluates only the power spectrum of the population output. 
Phase reorganization therefore becomes an available degree of freedom alongside single-neuron amplitude changes.

These mechanisms are not mutually exclusive. Different mechanisms can dominate different mode pairs within the same network, and, as Mechanism~C illustrates, the mechanism selected for a given mode pair can itself vary across training runs.

\subsection{Time constant diversity and functional differentiation}\label{sec:tau_discussion}

A notable finding of this study is that the relationship between learned time constants and functional differentiation depends qualitatively on the frequency band. 
In the $\gamma$ mode, the correlation between time constant and amplitude is very strong ($\rho_s^{\mathrm{amp}} \approx -0.82$): intrinsic time constants directly dictate functional role.
In the $\theta$ mode, by contrast, this correlation is weak ($\rho_s^{\mathrm{amp}} \approx -0.15$), and the recurrent dynamics play the dominant role.

This frequency dependence has an interesting correspondence with the theoretical observation of \citet{stern2023}, who showed that single-neuron time constant diversity alone does not necessarily produce time-scale separation under recurrent dynamics.
Our results are consistent with their observation in the low-frequency regime, that the generation of the $\theta$ rhythm does not depend on intrinsic time constants and is instead governed by the collective recurrent dynamics.
In the high-frequency regime, however, intrinsic time constants strongly constrain functional differentiation.
A natural interpretation is that high-frequency oscillations impose tight constraints on the intrinsic response speed of individual neurons, and these constraints cannot be fully compensated by recurrent coupling.

\citet{quax2020} reported that a two-layer RNN with adaptive time constants self-organizes a hierarchy of time scales across its layers.
Our results show that a single-layer RNN without explicit hierarchical architecture can likewise acquire, through training, a structure in which a continuous distribution of time constants maps systematically onto functional differentiation, suggesting that time-scale separation by time constants does not require hierarchical architecture as a prerequisite.

\subsection{Relation to prior work on rhythm-switching mechanisms}\label{sec:rhythm_discussion}

Among recent studies on rhythm generation in RNNs, \citet{zemlianova2024} trained an excitatory--inhibitory RNN on a rhythmic timing task and elucidated a mechanism in which a context input switches the state of the dynamical system to control the oscillation frequency.
Our baseline-shift mechanism (Mechanism~B) shares a similar control structure, in that the one-hot pulse moves the operating point into different attractor regions.
However, \citet{zemlianova2024} focus on modulation within a single tempo band (2--8\,Hz) and do not address the reorganization of the neuronal population that accompanies switching across a wide range of frequencies.

\citet{sussillo_barak2013} showed, in a sinusoidal generation task in which the frequency is specified continuously by an input signal, that the oscillation frequency of the trained RNN is controlled by an operating point on a low-dimensional slow manifold.
Our Mechanism~B is consistent with this picture: a change in baseline moves the operating point of the dynamical system and thereby sets the oscillation frequency.
Their analysis, however, focuses on the geometry of the macroscopic state space and does not examine how individual neurons functionally differentiate across frequency bands.
Our work complements theirs by showing that this macroscopic shift of the operating point is underpinned by selective, time-constant-dependent recruitment of neurons (Mechanism~A) and by reorganization of phase coherence (Mechanism~C), thereby describing the microscopic mechanisms of rhythm switching at single-neuron resolution.

\citet{pals2024} trained an RNN with low-rank recurrent connectivity on a working-memory task and showed that phase-locked limit cycles form between an autonomously generated internal oscillation and an external reference.
Our Mechanism~C also relies on phase relationships to control the population representation of a rhythm, but there is a key difference.
In \citet{pals2024}, phase locking is a macroscopic relationship between the network output and an external reference, whereas in Mechanism~C it is the reorganization of phase relationships \emph{among neurons within the network} that modulates the band-specific population output. 

\subsection{Transient versus tonic input and the role of multistability}\label{sec:input_discussion}

Because the switching cue in our task is a brief one-hot pulse rather than a sustained signal, the network must maintain each rhythm autonomously: after the $100$-step pulse the input is zero for the remaining several thousand steps of the interval, yet the designated rhythm persists. The four rhythms are therefore realized as coexisting, autonomously maintained states (stable attractors, or at least long-lived metastable states), and a rhythm switch corresponds to the transient pulse moving the state between them.
For the $\beta$ and $\gamma$ modes our fixed-point analysis makes this concrete: the mean activity of each of these modes sits near a distinct unstable fixed point whose linear instability lies within the corresponding band, and free-running the autonomous dynamics from that state reproduces the band (Sect.~\ref{sec:baseline_results}, Fig.~\ref{fig:baseline}). 
Of the three mechanisms, only the baseline shift (mechanism~B) is specific to this transient paradigm: its realization as selection among coexisting autonomous attractors would be unnecessary under a tonic input that holds the operating point in place, although the underlying principle---that the operating point sets the oscillation frequency---is general \citep{sussillo_barak2013, zemlianova2024}. The time-constant-dependent recruitment (mechanism~A) and the phase-interference route to band-power control (mechanism~C) reflect, respectively, the intrinsic response speed required for a given frequency and a property of the population-averaged readout, and are therefore not expected to depend on the input modality.

Finally, the transient, non-oscillatory one-hot cue was deliberate: it functions as a discrete contextual command that specifies which rhythm to produce---analogous to the rule and context inputs of cognitive-task RNN models \citep{yang2019task, zemlianova2024, sussillo_barak2013}---rather than a sensory signal carrying the rhythm itself, reflecting our focus on endogenous generation and switching rather than entrainment to an external drive.
We recognize that an area embedded in a larger circuit also receives oscillatory afferents, which would be the natural input for studying entrainment \citep{pals2024}; discrete command or gating signals and sustained oscillatory drive are both biologically relevant, probe complementary computations, and would be worth comparing within the same architecture in future work.

\subsection{Diversity of learned solutions}\label{sec:diversity_discussion}

A second important finding is that, for the same task and loss function, multiple internal structures can emerge as learned solutions (Sects.~\ref{sec:intermode_results}, \ref{sec:phase_results}).
The relative independence of the $\theta$-rhythm population is largely preserved across training runs. By contrast, both the population-sharing patterns among $\alpha$, $\beta$, and $\gamma$ and the microscopic mechanism of rhythm switching (amplitude-based vs.\ phase-interference-based) converge to different structures depending on training initialization.

This phenomenon is reminiscent of the concept of degeneracy in biological neural circuits \citep{edelman2001, marder2011} in which structurally distinct systems can realize the same function, which is considered a foundation for the robustness and adaptability of neural circuits.
Our results show that, even in RNNs trained by supervised learning, multiple structurally distinct solutions can implement the same input--output relationship, i.e., the trained network exhibits degeneracy.
Interestingly, \citet{pals2024} similarly reported that, in a single working-memory task, two qualitatively different solutions (phase-coding and rate-coding) emerge depending on initialization and training settings, suggesting that solution diversity in trained RNNs is not idiosyncratic to our setting but a more general property.

The asymmetry between the robustness of $\theta$ and the variability of the $\alpha$--$\beta$--$\gamma$ relations likely reflects differences in the computational constraints imposed by the task.
The $\theta$ rhythm is generated by broad participation across neurons spanning a wide range of time constants, qualitatively distinct from the concentrated, fast-neuron-driven generation of $\beta$ and $\gamma$. 
The intermediate-frequency $\alpha$ band can be served partly by the same fast neurons that drive $\beta$ and $\gamma$, allowing population sharing among $\alpha$, $\beta$, $\gamma$, while the $\theta$ population remains comparatively isolated.
By contrast, the $\beta$ (14--29\,Hz) and $\gamma$ (30--50\,Hz) bands are comparatively broad, and multiple solutions can be realized for the same band by placing the spectral peak at different frequencies within the band.
This is also a consequence of our loss function design (Eq.~\ref{eq:loss}), which evaluates only the band power ratio and leaves the detailed within-band spectral structure underdetermined.

Our findings also imply that analyses based on a single training run carry a risk of reaching overly specific conclusions about the computational mechanism of a network.
Statistical validation across multiple independent runs is essential for distinguishing robust features of the computational mechanism from run-specific artifacts.

\subsection{Implications for neuroscience}\label{sec:neuroscience_discussion}

The frequency-dependent functional differentiation discovered in this study stands in qualitative correspondence with the organization of rhythm generation in biological neural circuits.
In the cerebral cortex, low-frequency rhythms ($\theta$, $\alpha$) tend to synchronize globally through wide-ranging corticocortical and corticosubcortical projections, whereas high-frequency rhythms ($\beta$, $\gamma$) tend to be generated by a small number of neurons within local circuits \citep{buzsaki_draguhn2004}.
Our RNN exhibits the same pattern in which low-frequency rhythms are generated by distributed participation of many neurons, while high-frequency rhythms emerge from concentrated participation of a few high-amplitude neurons. This similarity arises without any anatomical constraints.

Furthermore, \citet{klausberger2008} reported that hippocampal interneurons include types such as O-LM cells that fire specifically with the $\theta$ rhythm alongside types such as basket cells that participate in both the $\theta$ and $\gamma$ rhythms.
In our RNN, a $\theta$-specific neuronal population coexists with populations shared among $\alpha$, $\beta$, and $\gamma$, and both modes of functional organization, namely specialization for a particular rhythm and involvement in multiple rhythms, emerge without any anatomical prior.

These correspondences should, however, be interpreted cautiously.
Our model does not obey Dale's law and lacks both cortical laminar structure and the distinction between interneuron subtypes.
In addition, the diversity of ion-channel compositions that set the effective time constants of biological neurons is not explicitly modeled.
The correspondences above are therefore suggestive of organizing principles for functional differentiation rather than direct identifications with specific biological mechanisms.

\subsection{Limitations and future directions}\label{sec:limitations}

This study has several limitations.
First, the network size ($N = 100$) is small, and how the ratio of dominant (high-amplitude) to broadly-participating neurons and the spatial structure of functional differentiation scale with network size remains to be determined.
Second, we did not incorporate Dale's law (the distinction between excitatory and inhibitory neurons) or detailed ion-channel properties, so it remains open whether similar functional differentiation arises in biologically constrained networks.
Third, the input structure is limited to simple one-hot pulses, with generalization to more complex inputs and to continuous contextual modulation left as future work.

These findings open several directions of broader neuroscientific interest. 
The three-mechanism framework provides a vocabulary for interpreting empirical neural recordings, offering candidate microscopic accounts of rhythm switching in local field potential and single-unit data. 
Our prediction that short-time-constant neurons are preferentially recruited for high-frequency rhythms can be tested against circuits with systematic time-constant variation across cell classes (e.g., parvalbumin-positive interneurons in cortical $\gamma$ generation). 
The degeneracy of learned solutions also suggests that observed inter-individual and inter-region variability in rhythm-generating circuits may reflect functional equivalence rather than imprecision.

In subsequent work we plan to pursue several methodological extensions: detailed characterization of state-space transitions through low-dimensional dynamics analysis, and extension to excitatory--inhibitory RNNs that obey Dale's law \citep{song2016training}.
Our linearized (Jacobian) analysis (Sect.~\ref{sec:connectivity_results}) could be complemented by other dynamical-systems tools---for instance dynamical mean-field theory for the statistics of the learned connectivity, or data-driven modal decompositions such as dynamic mode decomposition---to further probe the generative dynamics.
Together, these directions would deepen both the dynamical-systems account of the learned circuits and their connection to biology.

\backmatter

\section*{Statements and Declarations}

\noindent
\textbf{Funding.} This work was supported by Japan Society for the Promotion of Science (JSPS) KAKENHI grant numbers 23K11256 and 26K15000.

\noindent
\textbf{Competing interest.} The authors declare that they have no competing interests.

\noindent
\textbf{Ethics approval and consent to participate.} Not applicable.

\noindent
\textbf{Consent for publication.} Not applicable.

\noindent
\textbf{Data availability.} All data analyzed in this study were generated by the simulation and analysis code described in Sect.~\ref{sec:methods}. The code, together with scripts to reproduce the analyses and figures, is publicly available at \url{https://github.com/cncs-fit/rnn-rhythm} and archived at Zenodo (\url{https://doi.org/10.5281/zenodo.21900320}).

\noindent
\textbf{Materials availability.} Not applicable.

\noindent
\textbf{Code availability.} The code used to train the networks and to perform all analyses is publicly available at \url{https://github.com/cncs-fit/rnn-rhythm} and permanently archived at Zenodo (DOI: \href{https://doi.org/10.5281/zenodo.21900320}{10.5281/zenodo.21900320}).

\noindent
\textbf{Author contribution.} Conceptualization: YY; Methodology: SN, YY; Software: SN, YY; Formal analysis: SN, YY; Investigation: SN, YY; Writing -- original draft: SN, YY; Writing -- review and editing: YY; Supervision: YY; Project administration: YY; Funding acquisition: YY.

%\bibliography{refers}% common bib file

\begin{appendices}
% Per Biological Cybernetics guidelines, appendix figures continue the
% consecutive numbering of the main text (Fig. 8, 9, ...) rather than A1, B1, ...
\renewcommand{\thefigure}{\arabic{figure}}

\section{Distribution of $\bar{\kappa}$ across training runs}\label{secA1}
\setcounter{figure}{8}

\begin{figure*}[ht]
  \centering
  \includegraphics[width=\textwidth]{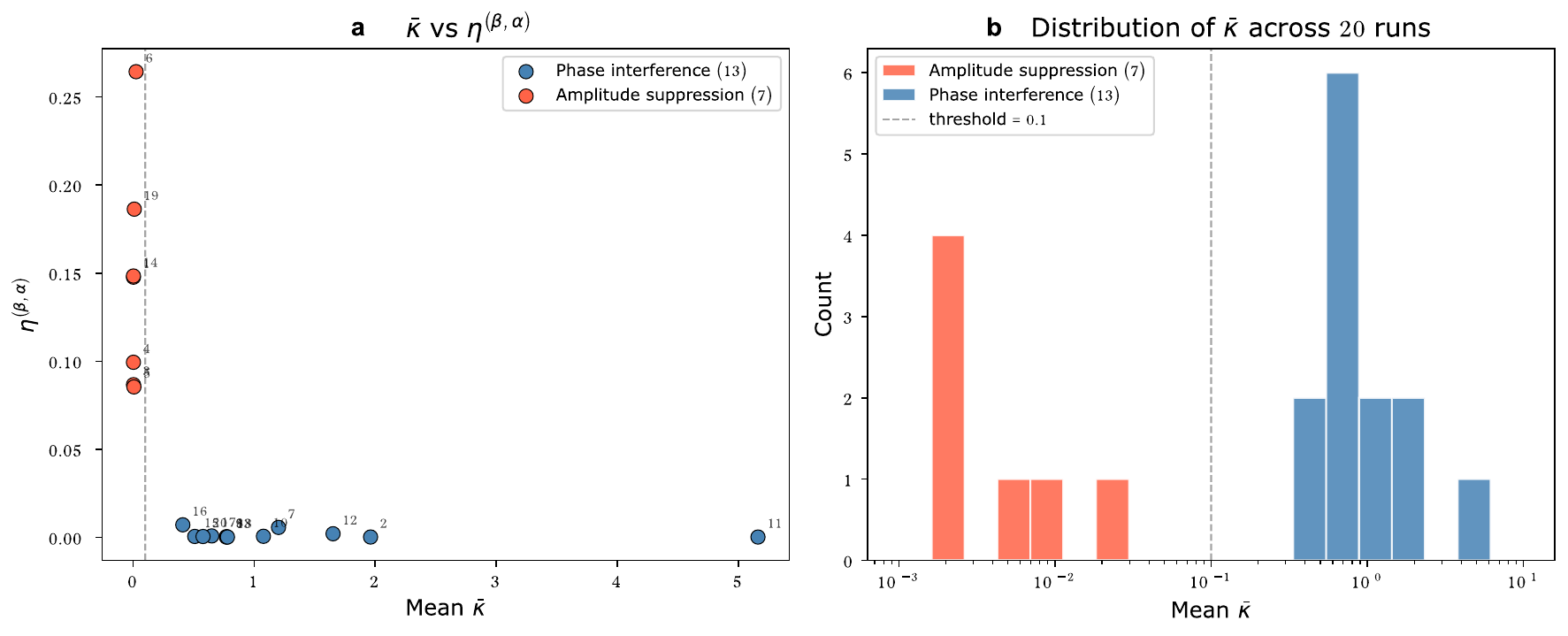}
  \caption{Distribution of the amplitude ratio $\bar{\kappa}$ across $n = 20$ runs and classification of switching mechanisms at the $\alpha \to \beta$ transition.
  $\bar{\kappa}$ is the neuron-averaged amplitude ratio $\kappa_i^{(\alpha;\,\alpha,\beta)}$ (Eq.~\ref{eq:amplitude_ratio}), quantifying how much of the single-neuron $\alpha$-band amplitude is retained in the $\beta$ mode.
  Runs with $\bar{\kappa} > 0.1$ (dashed line) are classified as the phase-interference type (blue; $n = 13$), in which individual neurons continue to oscillate in the $\alpha$ band during the $\beta$ mode but destructive inter-neuronal phase relations suppress the $\alpha$ component of the population output.
  Runs with $\bar{\kappa} \le 0.1$ are classified as the amplitude-suppression type (red; $n = 7$), in which the $\alpha$-band oscillation itself is attenuated at the single-neuron level.
  (a) Scatter plot of $\bar{\kappa}$ versus the $\alpha$-band coherence ratio $\eta^{(\beta,\alpha)}$ in the $\beta$ mode (Eq.~\ref{eq:coherence_ratio}); numbers annotate run IDs.
  (b) Histogram of $\bar{\kappa}$ (log-spaced bins). The distribution is bimodal, with the threshold $\bar{\kappa} = 0.1$ separating the two classes.}
  \label{fig:sup_kappa}
\end{figure*}

Figure~\ref{fig:sup_kappa} provides the full distribution of the neuron-averaged amplitude ratio $\bar{\kappa}$ (Eq.~\ref{eq:amplitude_ratio}) across all $n = 20$ networks, which underlies the classification of switching mechanisms at the $\alpha \to \beta$ transition (Sect.~\ref{sec:phase_results}). The distribution is clearly bimodal (panel~b): networks with $\bar{\kappa} > 0.1$, in which individual neurons keep oscillating in the $\alpha$ band during the $\beta$ mode (phase-interference type, $n = 13$), are well separated from those with $\bar{\kappa} \le 0.1$, in which the single-neuron oscillation itself is attenuated (amplitude-suppression type, $n = 7$). Plotting $\bar{\kappa}$ against the $\beta$-mode $\alpha$-band coherence ratio $\eta^{(\beta,\alpha)}$ (panel~a) shows that the same two groups are also separated along the coherence axis, confirming that they correspond to two distinct microscopic routes---phase reorganization versus single-neuron amplitude suppression---to the same population-level frequency shift.

\section{Causal silencing of time-constant-ranked subpopulations}\label{secB1}
\setcounter{figure}{9}

\begin{figure*}[ht]
  \centering
  \includegraphics[width=\textwidth]{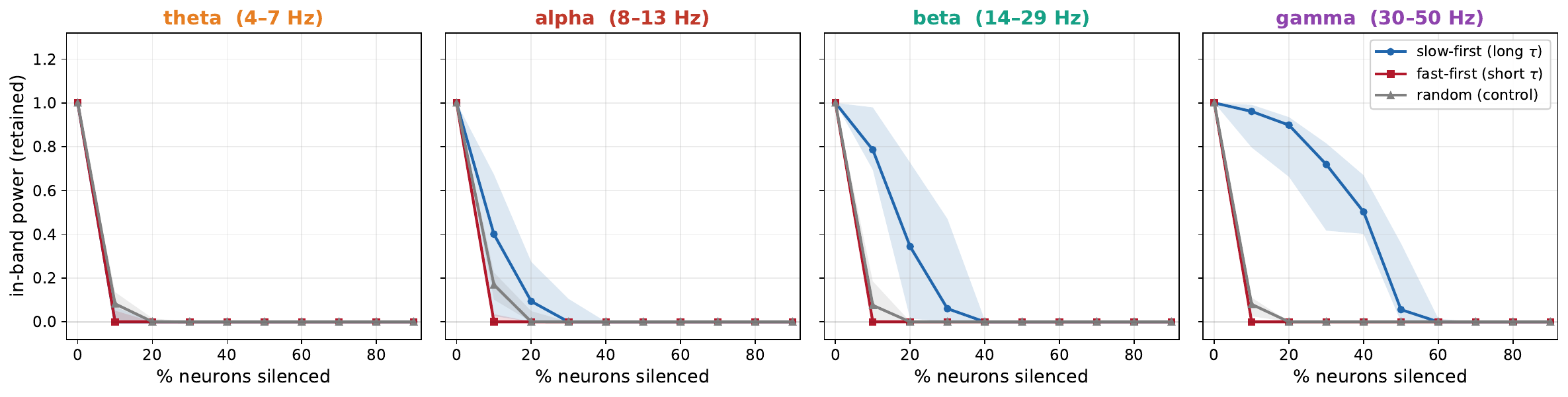}
  \caption{Causal silencing of neurons ranked by time constant (freeze method; all $n = 20$ networks).
  Each panel corresponds to one rhythm mode and plots the in-band power of the population output, retained relative to the intact network, as a function of the fraction of neurons silenced.
  Neurons are silenced from the longest time constant first (slow-first, blue), from the shortest first (fast-first, red), or in random order (control, gray).
  Lines and shaded bands show the median and interquartile range across the $20$ networks.
  In the freeze method, a silenced neuron's output is clamped to its per-mode mean firing rate, removing its oscillation while preserving its baseline contribution.
  For $\beta$ and $\gamma$, silencing the fastest neurons abolishes the rhythm while silencing the slowest neurons preserves it; for $\theta$ the rhythm is abolished regardless of which neurons are silenced, reflecting its distributed generation.}
  \label{fig:ablation}
\end{figure*}

\begin{figure*}[ht]
  \centering
  \includegraphics[width=0.75\textwidth]{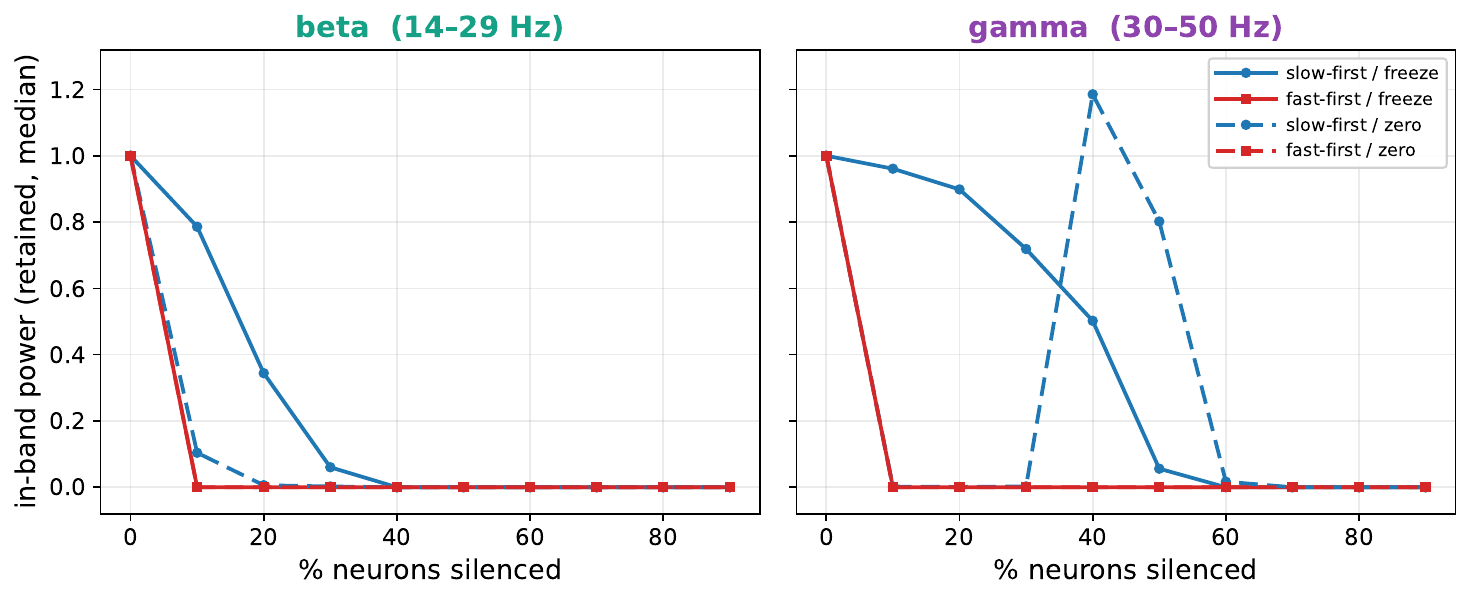}
  \caption{Zero-ablation versus freeze for the $\beta$ and $\gamma$ modes (median over $n = 20$ networks; slow-first in blue, fast-first in red; solid lines: freeze, dashed lines: zero-ablation).
  Fully removing (zero-ablating) the slowest neurons perturbs the high-frequency rhythms strongly and non-monotonically (unlike freezing), because it additionally disrupts the network-wide baseline that positions the operating point (mechanism~B, Sect.~\ref{sec:baseline_results}).
  Freezing preserves this baseline contribution and thereby isolates the oscillatory contribution, which is confined to the fast subpopulation.}
  \label{fig:ablation_supp}
\end{figure*}

Figure~\ref{fig:ablation} reports the causal silencing analysis (Sect.~\ref{sec:ablation}, \ref{sec:ablation_results}) using the freeze method for all four modes, and Fig.~\ref{fig:ablation_supp} compares the freeze and zero-ablation methods for the high-frequency modes.

\section{Inter-mode population sharing and the learned time-constant distribution}\label{secC1}
\setcounter{figure}{11}

\begin{figure*}[ht]
  \centering
  \includegraphics[width=\textwidth]{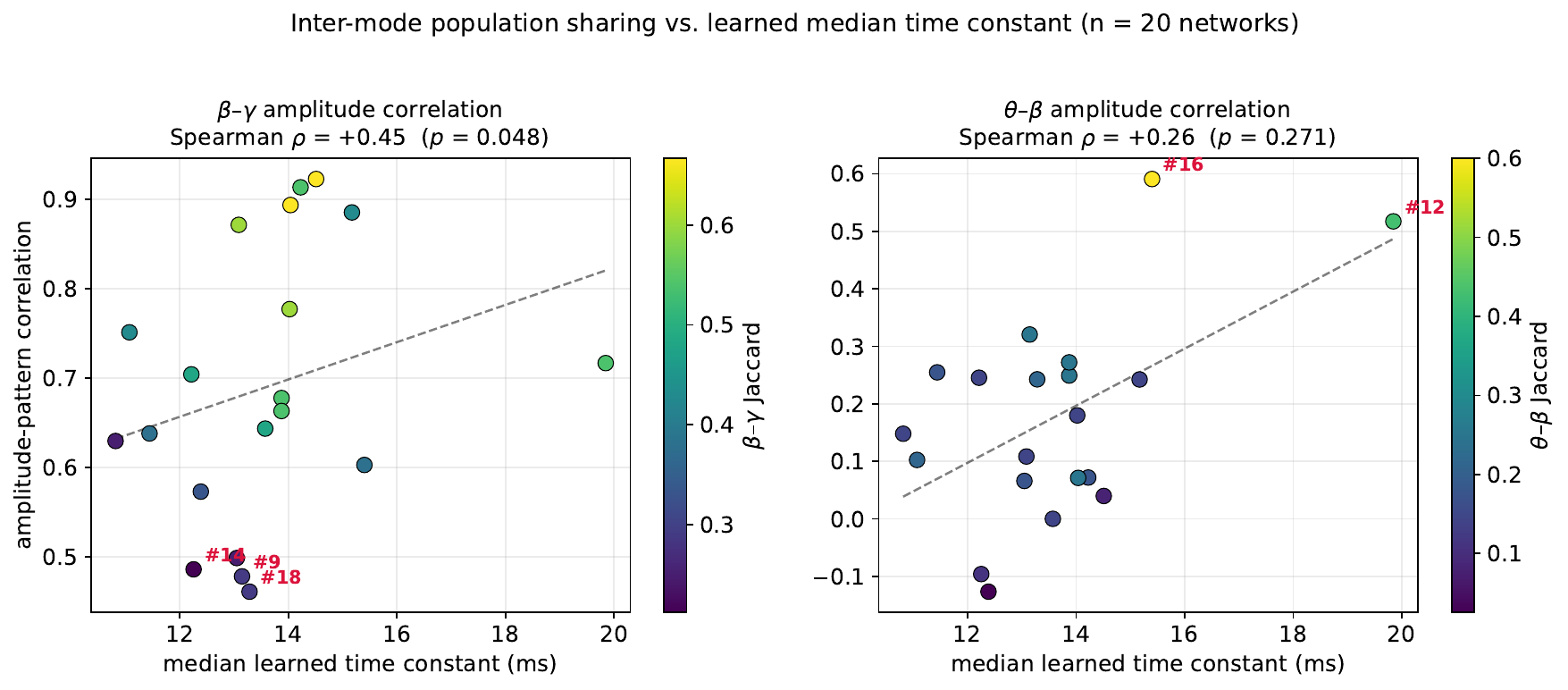}
  \caption{Inter-mode population sharing versus the learned median time constant across the $n = 20$ networks.
  Each point is one trained network; the horizontal axis is the median of its learned time-constant distribution.
  Left: $\beta$--$\gamma$ amplitude-pattern correlation (Eq.~\ref{eq:amp_corr}); right: $\theta$--$\beta$ correlation.
  Point color encodes the corresponding Jaccard coefficient of the active populations (Eq.~\ref{eq:amp_corr} and the Jaccard definition in Sect.~\ref{sec:intermode}), and the dashed line is a linear fit.
  Networks flagged as outliers (annotated by run ID; low $\beta$--$\gamma$ sharing on the left, high $\theta$--$\beta$ sharing on the right) tend to have distinctive median time constants, indicating that the run-to-run variability in inter-mode sharing is partly organized by the learned time-constant distribution.}
  \label{fig:outlier_tau}
\end{figure*}

To examine the run-to-run variability of inter-mode population sharing (Sect.~\ref{sec:intermode_results}, Fig.~\ref{fig:intermode}c), and in particular the networks that deviate from the typical pattern, we related the inter-mode amplitude-pattern correlations (Spearman; Eq.~\ref{eq:amp_corr}) to the learned time-constant distribution of each network (Fig.~\ref{fig:outlier_tau}). The $\beta$--$\gamma$ correlation increases with the median learned time constant ($\rho = 0.45$, $p = 0.048$): networks with a lower median time constant separate $\beta$ and $\gamma$ into distinct fast subpopulations (the low-sharing outliers), whereas networks with a higher median time constant share the fast pool across both bands. The $\theta$--$\beta$ correlation shows a similar but weaker and non-significant trend ($\rho = 0.26$, $p = 0.27$); the two networks with the highest $\theta$--$\beta$ sharing nonetheless had among the longest median time constants (runs~12 and~16, median $\tau = 19.8$ and $15.4$\,ms vs.\ a typical $\approx 13$\,ms). That the median-time-constant dependence is significant for $\beta$--$\gamma$ but not for $\theta$--$\beta$ is consistent with the interpretation above: the size of the fast pool governs sharing among the higher-frequency bands, whereas $\theta$ is generated by the broadly distributed population and therefore does not follow the same rule. The Jaccard coefficient of the active populations (color in Fig.~\ref{fig:outlier_tau}) co-varies with the amplitude correlation, indicating that the two measures reflect closely related structure.

\section{Robustness to time-constant initialization}\label{secD1}
\setcounter{figure}{12}

\begin{figure*}[ht]
  \centering
  \includegraphics[width=\textwidth]{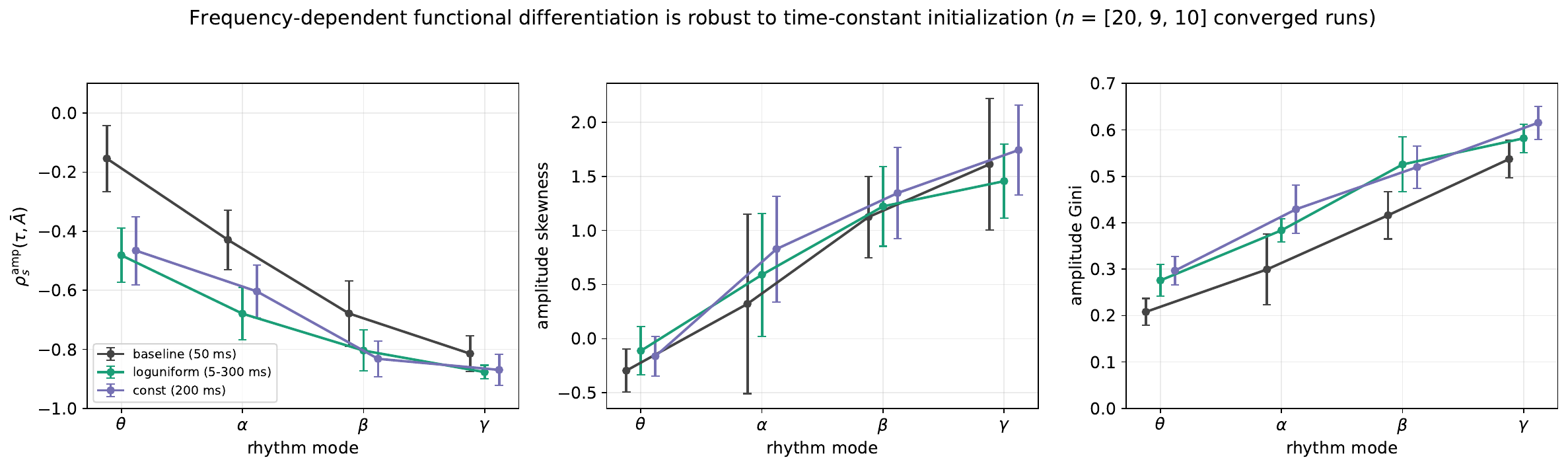}
  \caption{Frequency-dependent functional differentiation under three time-constant initializations: baseline (all $\tau = 50$\,ms; the main $n = 20$ networks), per-neuron log-uniform over $5$--$300$\,ms ($n = 9$ converged), and uniform $\tau = 200$\,ms ($n = 10$). Points and error bars are mean $\pm$ SD across networks.
  Left: Spearman correlation between the learned time constant and the matched-mode amplitude, $\rho_s^{\mathrm{amp}}$, per mode. Middle: amplitude skewness. Right: amplitude Gini coefficient.
  The monotonic strengthening from $\theta$ to $\gamma$ is present in all three schemes and the high-frequency bands nearly coincide; the schemes differ mainly at the low frequencies, where seeding slow neurons strengthens the $\theta$/$\alpha$ correlations.}
  \label{fig:reinit_metrics}
\end{figure*}

\begin{figure*}[ht]
  \centering
  \includegraphics[width=0.7\textwidth]{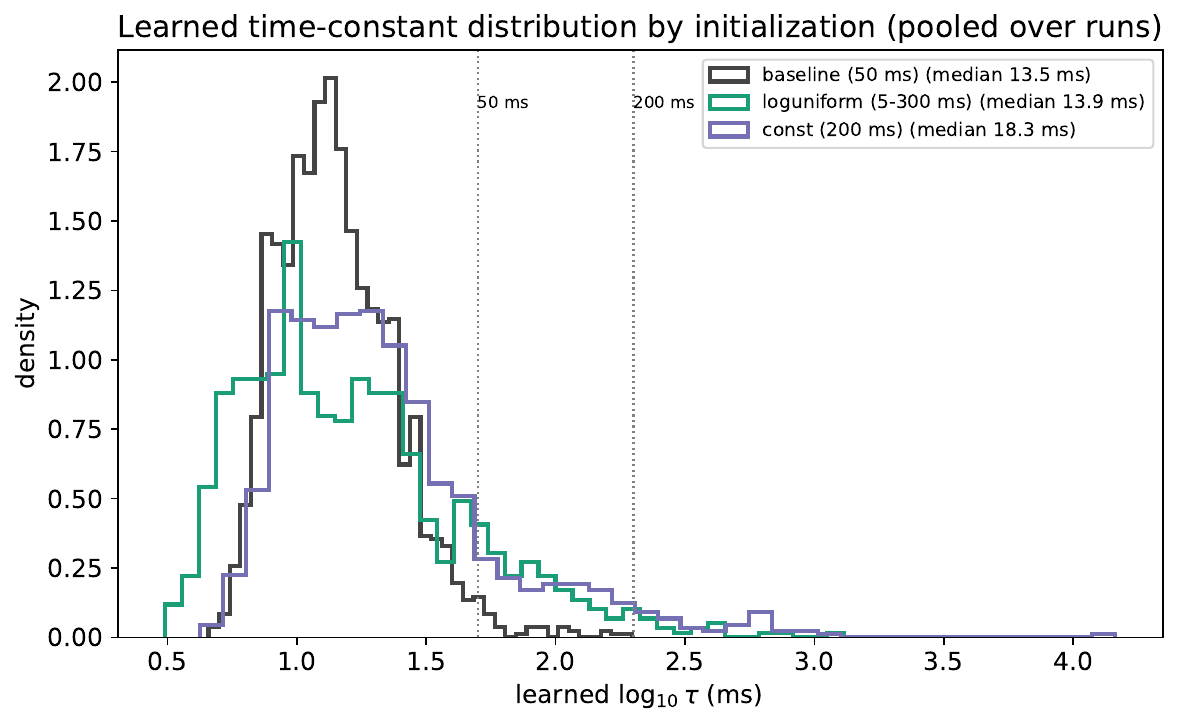}
  \caption{Distribution of the learned time constants (pooled over networks, log scale) for the three initialization schemes; dotted lines mark the $50$\,ms and $200$\,ms initialization values. Regardless of the initialization the bulk of the time constants converges to short values (medians $13.5$, $13.9$, $18.3$\,ms), while the broader/longer initializations leave a heavier tail of retained slow neurons.}
  \label{fig:reinit_tau}
\end{figure*}

Figures~\ref{fig:reinit_metrics} and~\ref{fig:reinit_tau} report the time-constant-initialization control of Sect.~\ref{sec:reinit_results} in full. In all three schemes (baseline $50$\,ms; per-neuron log-uniform $5$--$300$\,ms; uniform $200$\,ms) the amplitude--time-constant correlation strengthens monotonically from $\theta$ to $\gamma$ and the amplitude concentration (Gini) increases with frequency (Fig.~\ref{fig:reinit_metrics}); the high-frequency bands are essentially indistinguishable across schemes ($\rho_s^{\mathrm{amp}}(\gamma) = -0.81$ to $-0.88$). Regardless of the initialization the networks drive the bulk of the time constants to short values (median $13.5$, $13.9$, $18.3$\,ms; Fig.~\ref{fig:reinit_tau}), while the broader and longer initializations retain a heavier tail of slow neurons. The initialization does, however, modulate the \emph{low-frequency} correlations: when slow neurons are available, the $\theta$ and $\alpha$ amplitude--time-constant correlations become substantially stronger ($\rho_s^{\mathrm{amp}}(\theta) = -0.48$ and $-0.47$ for the log-uniform and $200$\,ms schemes, versus $-0.15$ at baseline). This indicates that the network recruits intrinsically slow neurons for the low-frequency rhythms when they are available but generates the same rhythms through recurrent dynamics when they are not (Sect.~\ref{sec:tau_discussion}, \ref{sec:connectivity_results}); the high-frequency differentiation thus robustly requires short intrinsic time constants, whereas low-frequency generation can flexibly use either substrate.

\end{appendices}

\end{document}